\documentclass[twocolumn,showpacs,amsmath,prb,superscriptaddress]{revtex4-1}
\usepackage{epsfig,psfrag}
 \usepackage{amsbsy}
\usepackage{amsfonts}
\usepackage{wasysym}
\usepackage{bbm}
\usepackage{tabularx}
\usepackage{euscript}
\usepackage{amsfonts}
\usepackage{exscale}
 \usepackage{slashed}
\usepackage{comment}

\usepackage{subcaption}

\usepackage{amsmath,amssymb,wasysym}
\usepackage{graphicx}
\usepackage{dcolumn}
\usepackage{bm}
\usepackage{braket}
\usepackage{verbatim}
\usepackage{float}
\usepackage{bbold}
\usepackage{color}
\usepackage{xcolor}
\usepackage{relsize}
\usepackage{amsthm}
\usepackage{enumerate}
\usepackage{soul,xcolor}
\usepackage[T1]{fontenc}
\usepackage[colorlinks=true ,urlcolor=blue,urlbordercolor={0 1 1}]{hyperref}

\newcommand{\beq}{\begin{eqnarray}}
\newcommand{\eeq}{\end{eqnarray}}

\newcommand{\bsp}{\begin{split}}
\newcommand{\esp}{\end{split}}

\newcommand{\be}{\begin{equation}}
\newcommand{\ee}{\end{equation}}

\begin{document}

\setstcolor{red}

\title{Pair-density-wave superconductor from doping Haldane chain and rung-singlet ladder}
\author{Ya-Hui Zhang}
\affiliation{Department of Physics, Harvard University, Cambridge, Massachusetts 02138, USA}
\affiliation{Department of Physics and Astronomy, Johns Hopkins University, Baltimore, Maryland 21218, USA}

\author{Ashvin Vishwanath}
\affiliation{Department of Physics, Harvard University, Cambridge, Massachusetts 02138, USA}

\date{\today}

\begin{abstract}
We report the numerical discovery of a pair-density-wave (PDW) superconductor from doping either (i) a spin-one Haldane chain or (ii) a two-leg ladder in the rung singlet phase in which the doped charges occupy a single leg.  We model these systems using a generalized Kondo model. The itinerant electrons are correlated and described by the $t-J$ model, and are further coupled to a spin $1/2$ Heisenberg model through the Kondo coupling $J_K$. When the density of electrons $x$ is one, the Mott insulator is in a Haldane phase or in a rung singlet phase depending on whether $J_K$ is negative or positive. Upon doping, a pair-density-wave with $\mathbf Q=\pi$ can emerge for both signs of $J_K$. In the $J_K \rightarrow -\infty$ limit, the model reduces to the recently proposed type II t-J model and we observes a continuous transition between the PDW superconductor and an unconventional Luttinger liquid phase with doping. We also identify a composite order parameter for the superconductor, which can be understood as a Cooper pair formed by two nearby fermionic spin-polarons. Our model and the predicted PDW phase can be experimentally realized by doping  S=1 chains formed by Ni$^{2+}$ in a solid state system or a two-leg ladder of fermionic cold atoms with a potential bias between legs, which preferentially   dopes carriers into a single leg.  
\end{abstract}
\maketitle{}

\textbf{Introduction} The possibility of a high Tc superconductor emerging on doping a spin-{\em one-half} Mott insulator has been intensively studied in the last several decades\cite{lee2006doping}. However, the fate of doping a spin-{\em one} Mott insulator is not well explored so far. Here we take the first steps to doping a spin-one magnet by focusing on one dimension.  The one dimensional spin-one antiferromagnet is well known to be in the Haldane phase\cite{haldane1983nonlinear,haldane1983nonlinear,affleck1989quantum,affleck2004rigorous}, which is a classical example of symmetry protected topological ordered phase with edge modes\cite{gu2009tensor,pollmann2010entanglement}. 
 
Here however we will be interested in closing the charge gap by doping.  Physically the S=1 moments arise from electrons occupying two strongly correlated orbitals coupled together by a ferromagnetic Hund's coupling $J_H$\cite{fazekas1999lecture}.  Upon doping, holes enter one orbital due to crystal field splitting while the other orbital remains Mott localized\cite{zhang2020type,zhang2021fractional}.  It is the fate of this doped system that we are interested in here. Going beyond the large Hund's coupling regime, we consider an extended range of inter-orbital Kondo coupling $J_K=-J_H<0$. We also discuss the case with a positive $J_K$, which can naturally be realized in bilayer optical lattices\cite{gall2021competing,sompet2021realising} with a potential difference.   We  will report numerical discoveries of a pair-density-wave superconducting phase for both signs of $J_K$. A PDW superconductor has Cooper pair condensed at a non-zero momentum\cite{agterberg2020physics} so that the effect of translation symmetry combined with a phase rotation, leaves the order parameter invariant.

Various other  models have already been  studied  for doping a S=1 chain.  (I) In the first class of model, the $S=1$ moment in the undoped insulator is formed by one single electron with $S=1$\cite{ning2020topological} or three flavors\cite{keselman2018one}. These are not realistic models of $S=1$ in a solid state system where the $S=1 $ is  built from single electrons which carry only $S=\frac{1}{2}$. (II) In Ref~\onlinecite{zhu2018pairing,jiang2018symmetry},  the authors assume that the $S=1$ moment is formed by two spin 1/2 electrons with ferromagnetic inter-orbital (inter-leg) coupling $J_H$. However, in that model, there is no repulsion between the two orbitals and a tightly bound on-site Cooper pair is doped into the system while the single electron excitation is gapped  with binding energy $J_H$. We believe this class of model also does not capture the physics in real systems.

In our model, the $S=1$ moment in the undoped insulator is formed by two electrons on two orbitals. Then, we dope only one orbital with holes, while the other orbital is still singly occupied and just provides spin 1/2 local moments.  Thus we obtain a Kondo like model, with itinerant electron in a {\bf C} layer, which couples to $S=1/2$ local moments in an {\bf S} layer with a Kondo coupling $J_K$.  We label the density of the C layer as $x$. When $x=1$, we have a Mott insulator in a Haldane chain phase or in a rung singlet depending on the sign of $J_K$.  $J_K<0$ arises from Hund's coupling in 1D chain where the two legs represent the two  orbitals. In the $J_K\rightarrow -\infty$ limit, the model reduces to a new kind of t-J model dubbed as type II t-J model by us in a previous paper\cite{zhang2020type}. This type II t-J model interpolates between a spin 1/2 Mott insulator and a spin one Mott insulator by tuning the density $x$ from 0 to 1.

 The model with $J_K>0$ can be realized by doping a two-leg ladder with a potential bias. Unlike previous studies which dope both legs\cite{giamarchi2003quantum}, in our case only one leg is doped due to orbital or layer selective Mott localization caused by potential difference.  Finally, we note that a PDW phase with $\mathbf Q=\pi$ was previously reported in a Kondo-Heisenberg model with the Heisenberg coupling $J=2t$\cite{berg2010pair,cho2014topological,may2020topology}. Such a model resembles our generalized Kondo model with $J_K>0$. However,  there are also some essential differences. In our model, the C layer itself is also strongly correlated and is described by a t-J model. This stabilizes the PDW phase which is now realized  at a realistic parameter value of $J=0.5t$, which could be potentially realized in  two-leg ladder Fermi gas optical lattices.   Recently, a  cold atoms setup consisting of a bilayer optical lattice in the rung singlet state, with an interlayer potential difference, was modeled in Ref.~\onlinecite{bohrdt2021strong,bohrdt2021exploration}. In those works, a metastable configuration is studied, in which both layers (or both legs in 1D) are doped with  an equal density of charge carriers while the inter-layer tunneling is assumed to be zero due to a large potential difference. In contrast, our setup considers the ground state configuration where all the holes preferentially occupy one layer.  This leads to significant differences including the emergence of PDWs in our model.

\begin{figure}[ht]
\centering
\includegraphics[width=0.5\textwidth]{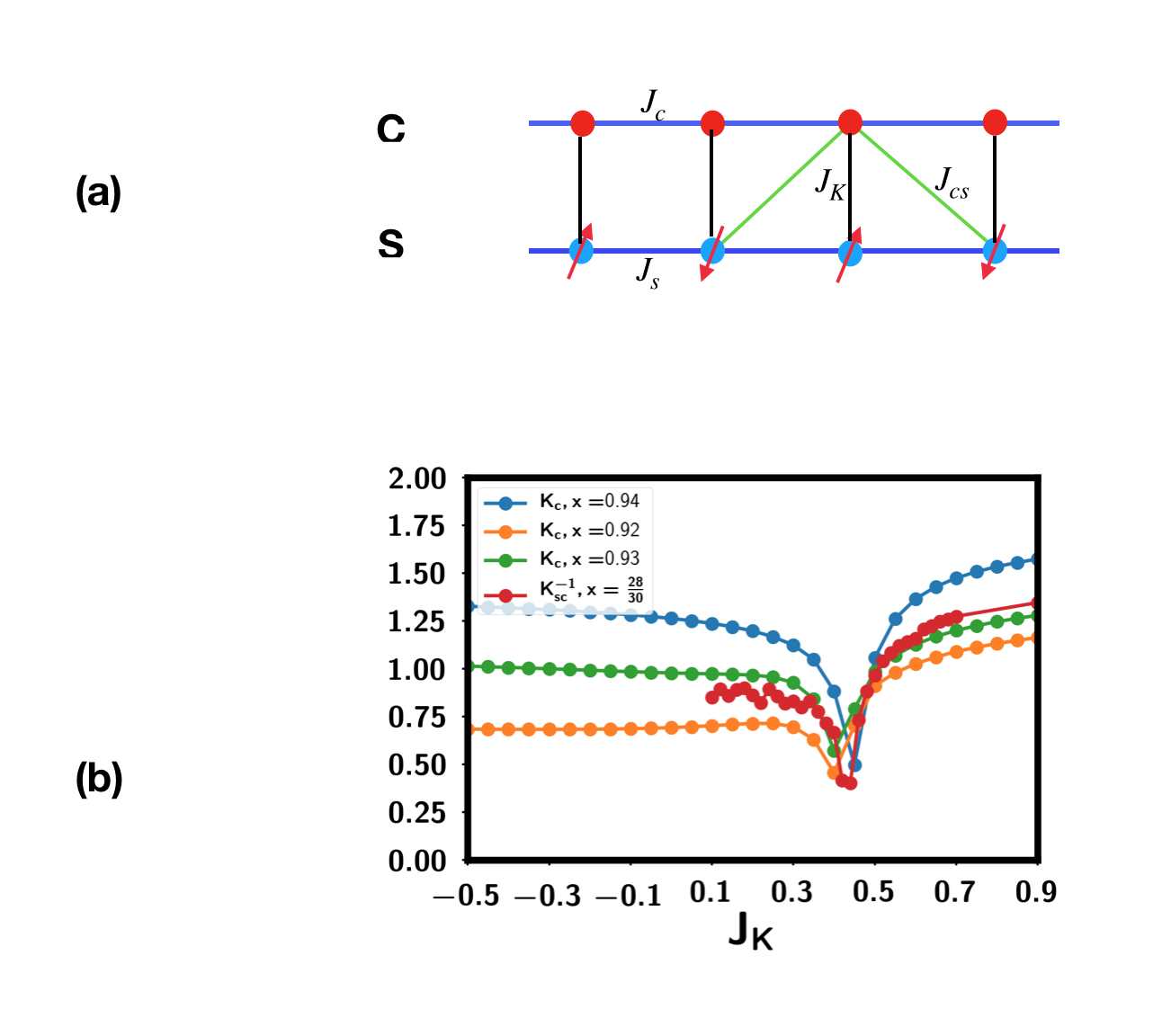}
\caption{(a) Illustration of the generalized Kondo model defined in Eq.~\ref{eq:kondo_t_J}. The C leg hosts the correlated itinerant electrons, while the S leg corresponds to local spin $1/2$ moment coming from Mott localization of another, lower energy, orbital. $J_c,J_s, J_{cs}$ are anti-ferromagnetic super-exchange terms. $J_K$ is the on-site Kondo coupling. (b) The inverse of the power law decay exponent for the pairing correlation function $K_{sc}^{-1}$ and the Luttinger parameter $K_c$ as a function of $J_K$. We used $J_{cs}=0.25$.  $K_{sc}$ is fit from infinite DMRG result at $x=\frac{28}{30}\approx 0.933$. $K_c$ is fit from finite DMRG results at several fillings at size $L_x=100$. We find $K_{sc}^{-1}\approx K_c$ evaluated at $x=0.93$. Luttinger parameter $K_c$ has a dip at $J_K=J_K^0\approx 0.45$, separating the PDW phase into two domes. }
\label{fig:kdon_hamiltonian}
\end{figure}

\textbf{Generalized Kondo model} To model a spin-one Mott insulator in solid state system, we can start from a two-orbital Hubbard model with Hund's coupling $J_H$ and  a crystal field splitting between the two orbitals.  As shown in the supplementary, in the end we obtain a generalized Kondo model as shown in Fig.~\ref{fig:kdon_hamiltonian}(a).

\begin{align}
H&=-t \sum_{ij;\sigma}P c^\dagger_{i;\sigma} c_{j;\sigma} P+J_c \sum_{\langle ij \rangle}\vec{S}_{i;c} \cdot \vec{S}_{j;c}+J_s \sum_{\langle ij \rangle}\vec{S}_{i} \cdot \vec{S}_{j}\notag\\
&~~+J_K \sum_i \vec{S}_{i;c}\cdot \vec{S}_i+J_{cs} \sum_{ij}\vec{S}_{i;c}\cdot \vec{S}_j
\label{eq:kondo_t_J}
\end{align}
where $J_K=-J_H$ is the ferromagnetic on-site Hund's coupling.  The first line is a conventional spin $1/2$ t-J model with $P$ as the projection operator to forbid double occupancy of the itinerant electron.  $\vec{S}_i$ represents the local spin $1/2$ moment.  $\vec{S}_{i;c}=\frac{1}{2} c^\dagger_{i;\sigma}\vec{\sigma}_{\sigma \sigma'}c_{i;\sigma'}$ is the spin of the electron in the C layer. We define the density of electron to be $x$ per site. When $x=1$, we have an insulator which is in either a Haldane  phase or in a rung-singlet phase depending on the sign of  $J_K-2J_{cs}$. Indeed we will see (Figure.~\ref{fig:kdon_hamiltonian}(b)) that various physical quantities defined below, dip in the vicinity of $J_K\approx 0.5$ (when $J_{cs}=0.25$), consistent with this expectation. We emphasize that finite super-exchange couplings $J_{s},J_c$ are important. Without them, the ground state is likely to be in a ferromagnetic phase due to the double exchange mechanism for x<1\cite{kagan1999double}. $J_{cs}$ is not necessary but its existence will enhance the PDW phase.  We will always use $t=1,J_c=J_s=0.5$ while varying $J_K$, $J_{cs}$ and $x$ in our calculation.

\begin{figure}[ht]
\centering
\includegraphics[width=0.45\textwidth]{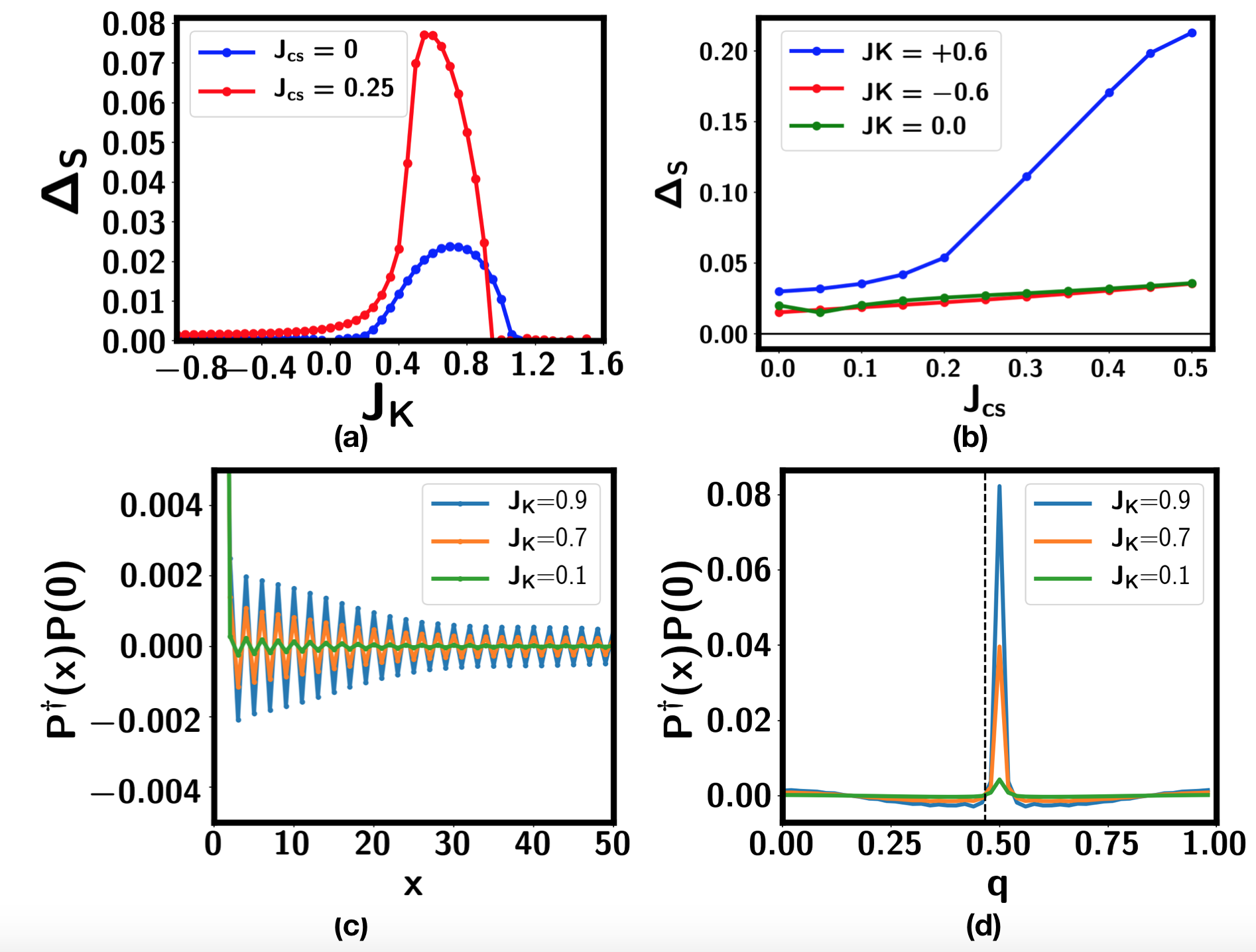}
\caption{Evidence for PDW phase in the generalized Kondo model defined in Eq.~\ref{eq:kondo_t_J}.  (a) Spin gap $\Delta_S(L_x=\infty)$ as a function of $J_K$ for $J_{cs}=0, \,0.25$ at the doping $x=0.9$ in units of $t=1$. The value is obtained from extrapolation of finite size results with $L_x=80,100,120$.  There is still a finite spin gap in the negative $J_K$ regime if one zooms in\cite{SM}.  The spin gap closes at $J_K^c \approx 1.05, \,0.95$ for $J_{cs}=0, \,0.25$. (b) Spin gap as function of $J_{cs}$ at $x=0.96$ using system size $L_x=100$ for fixed $J_K=-0.6,0,+0.6$. (c) Pairing-pairing correlation function in real space from infinite DMRG. We use $J_{cs}=0.25$ and $x=\frac{28}{30}\approx 0.933$ with unit cell size $L=30$.  Here $P(x)=\epsilon_{\sigma \sigma'}c_{\sigma}(x) c_{\sigma'}(x+1)$ is the spin-singlet Cooper pair on a nearest neighbor bond. (d) The Fourier transformation of the pairing-pairing correlation, peaked at $q=\pi$. Here $q$ is in unit of $2\pi$ and the dashed line labels $2k_F=\frac{x}{2} \times 2\pi$. }
\label{fig:PDW_kondo}
\end{figure}

\textbf{PDW superconductor} In Fig.~\ref{fig:PDW_kondo} we show evidences for a Luther-Emery liquid with  quasi long range PDW order parameter. In Fig.~\ref{fig:PDW_kondo}(a) we show a finite spin gap $\Delta_S$ when $J_K\in (-\infty,J_K^c)$. We note that a non-zero $J_{cs}$ can enhance the spin gap (see Fig.~\ref{fig:PDW_kondo}(b)).   When $J_K>J_K^c$ ($J_K^c \approx 0.95$ for $J_{cs}=0.25$), the ground state is in a Luttinger liquid phase with zero spin gap. Inside the spin gap phase, there is power law decay for the correlation function of the spin-singlet pairing order parameter: $\langle P^\dagger(x) P(0) \rangle \sim (-1)^x\frac{1}{x^{K_{sc}}}$, shown in Fig.~\ref{fig:PDW_kondo}(c). The oscillations $(-1)^x$ implies that the order parameter has a momentum $\mathbf Q=\pi$, hence the phase is a PDW superconductor.   We extract  $K_{sc}$ and the Luttinger parameter $K_c$ as shown in Fig.~\ref{fig:kdon_hamiltonian}(b). We find $K_c=K_{sc}^{-1}$ as expected for a Luther-Emery liquid. $K_c$ can be larger than 1, indicating slower decay of the superconductor (SC) order than the charge-density-wave (CDW) order.  However, when varying $J_K$ at fixed $J_{cs}=0.25$, $K_c$ has a dip at $J_K^0\approx 0.45$, which separates the PDW phase into two domes.

It turns out the inter-leg  spin-spin correlation changes from ferromagnetic to antiferromagnetic precisely at $J_K^0$.  We define a rung spin-correlator $V_i=\langle \vec{S}_{i;c}\cdot \vec{S}_{i}\rangle+\frac{1}{4}$ to characterize the inter-leg spin-spin correlation. $V$ changes sign around $J_K^0$, as shown in Fig.~\ref{fig:polaron}(a).   As shown in the supplementary\cite{SM}, there is a rapid crossover but no phase transition, at $J_K^0$. At $J_K^c\approx 0.95$, there is a small jump of $V$, suggesting a first order transition between the PDW phase and the Luttinger liquid phase with zero spin gap.

\begin{figure}[ht]
\centering{}
\includegraphics[width=0.5\textwidth]{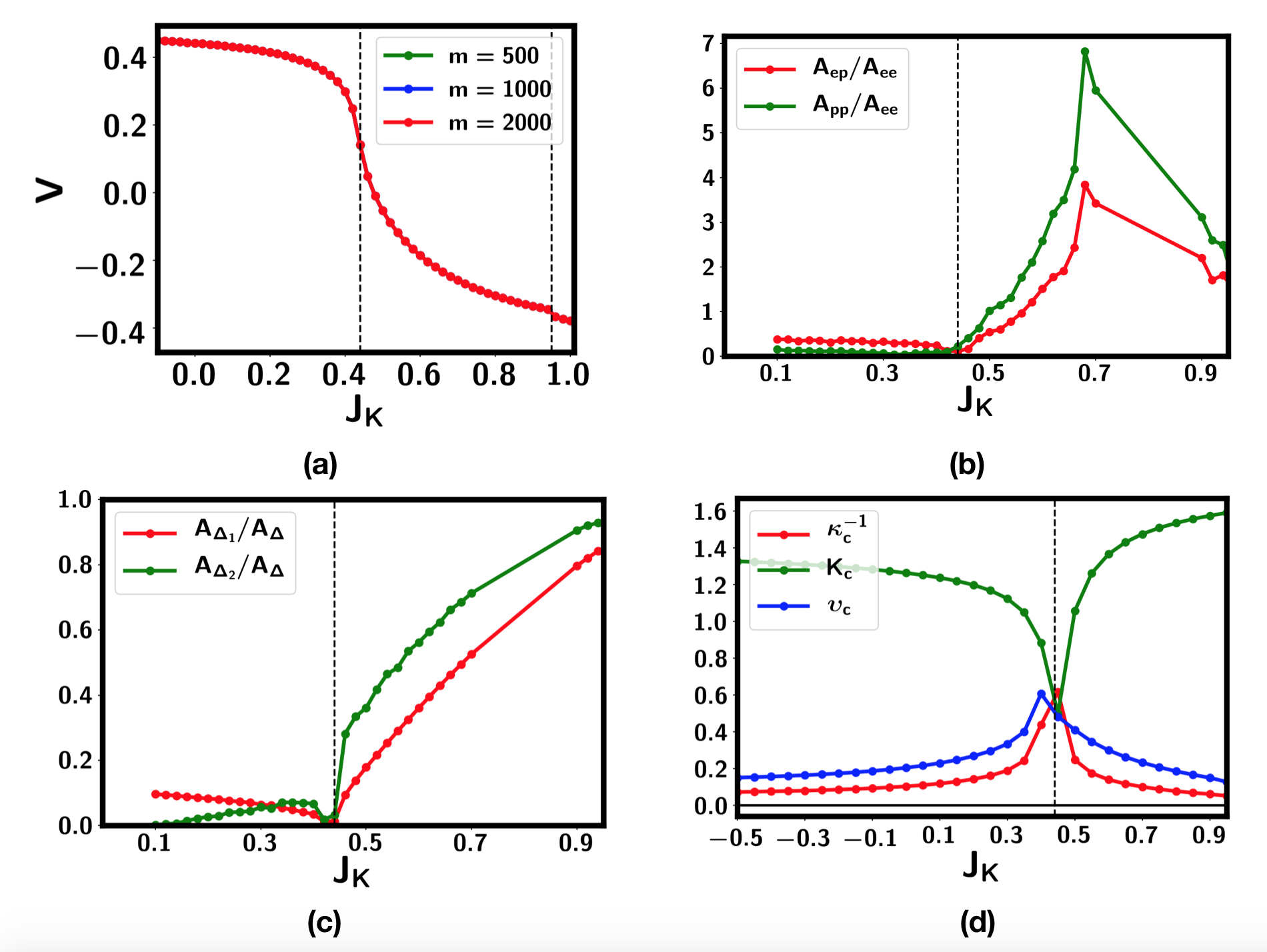}
\caption{Evolution with $J_K$ at fixed $J_{cs}=0.25$. (a) Rung spin-correlator $V=\langle \vec{S}_{i;c}\cdot \vec{S}_{i}\rangle+\frac{1}{4}$ from infinite DMRG at $x=\frac{28}{30}\approx 0.933$ with a unit cell size of $L=30$. The two dashed lines are at $J_K^0=0.44$ and $J_K^c=0.95$. Here $m$ is the bond dimension. (b) Amplitudes for Green functions of electron or polaron.  The vertical dashed line is at $J_K^0=0.44$.  (c) Amplitudes  for pairing-pairing correlation of composite Cooper pairs.  The vertical dashed line is at $J_K^0=0.44$, on either side of which the composite operator amplitudes grow. The same parameters are used as in (a)(b). (d) Inverse charge compressibility $\kappa_c^{-1}$, Luttinger parameter $K_c$, Fermi velocity of the charge mode $\upsilon_c$  at $x=0.94$ from finite DMRG with system size $L_x=100$.  We have used $\kappa_c=\frac{\pi}{2}\frac{K_c}{\upsilon_c}$ to extrapolate $\upsilon_c$. The vertical dashed line is at $J_K^0=0.45$, which separates the two PDW domes.}
\label{fig:polaron}
\end{figure}

We also point out the existence of spin-polarons at low energy when moving away from $J_K^0$ towards both sides. The spin-polaron is a bound state of electron in the C layer and spin operator in the S layer:

\begin{equation}
	\tilde c_{i;\sigma}=\frac{1}{2}(\vec{S}_i \cdot \vec{\sigma}_{\sigma \sigma'})c_{i;\sigma'}
\end{equation}
where $\vec \sigma$ is the Pauli matrix and $\vec{S}$ is the spin operator of the S layer.  This composite operator $\tilde c_{\sigma}$ has the same quantum number as the microscopic electron operator $c_{\sigma}$.

It can be easily shown that $c^\dagger_{i;\sigma}\tilde c_{i;\sigma}=\vec{S}_{i;c} \cdot \vec{S}_{i}$. Thus for either sign of $\langle \vec{S}_{i;c} \cdot \vec{S}_{i} \rangle$, there is a hybridization between the electron and the spin polaron. To characterize this hybridization, we define several different Green functions: the usual electron Green function $G_{\sigma;ee}(x,y)=\langle c^\dagger_{\sigma}(x) c_{\sigma}(y) \rangle$; electron-polaron Green  function $G_{\sigma;ep}(x,y)=\langle c^\dagger_{\sigma}(x) \tilde c_{\sigma}(y) \rangle_c$ and polaron-polaron Green  function $G_{\sigma;pp}(x,y)=\langle \tilde c^\dagger_{\sigma}(x) \tilde c_{\sigma}(y) \rangle_c$. Here we have subtracted the average values  so that $G_{\sigma;ep}(x,y)=G_{\sigma;pp}(x,y)=0$ in the decoupled limit $J_K=J_{cs}=0$. For each type $\alpha=$ ee, ep, pp, we find that $G_{\sigma;\alpha}(x,0)=A_{\alpha}e^{-\frac{x}{\xi}}$ in the PDW phase with the same exponent $\xi$. In Fig.~\ref{fig:polaron}(b),  we show that $\frac{A_{ep}}{A_{ee}},\frac{A_{pp}}{A_{ee}}$ vanish at $J_K^0$. Across $J_K^0$, the inter-layer spin-spin correlation changes sign and the mixture between the polaron and the electron vanishes.  This coincides with the dip of $K_c$, strongly suggesting that the existence of polaron is crucial for a large $K_c$, presumably from effective attractive interaction.  Especially in the $J_K^0<J_K<J_K^c$ regime, the ratio of amplitudes $\frac{A_{pp}}{A_{ee}}>>1$, indicating dominance of the polaron at low energy. This is also the regime where the PDW superconductor is strongest and the decay of the pairing correlation is the slowest.

 With the spin-polaron, we can also define the composite Cooper pair of an electron and a polaron, or a Cooper pair of two polarons.  We can define the usual spin-singlet Cooper pair $\Delta(x)=\epsilon_{\sigma \sigma'}c_{\sigma}(x)c_{\sigma'}(x+1)$ and two composite Cooper pairs: $\Delta_1(x)=\frac{1}{3} \big(\vec{\Delta}_T(x) \cdot (\vec{S}(x)-\vec{S}(x+1))\big)$ and $\Delta_2(x)=\Delta(x) \vec{S}(x)\cdot \vec{S}(x+1)$. In the above $\vec{\Delta}_T(x)$ is the spin-triplet Cooper pair on the  nearest neighbor bond.    In the supplementary\cite{SM} we show that $\Delta_1$ is the Cooper pair between electron and polaron while $\Delta_2$ is  from the Cooper pair of two polarons. We expect the existence of all  these three Cooper pairs at low energy. Indeed we find that $\langle \Delta^\dagger_{\alpha}(x)\Delta_{\alpha}(0)=A_{\alpha}\frac{(-1)^x}{x^{K_{SC}}}$ for $\alpha=\Delta,\Delta_1,\Delta_2$ with the same exponent $K_{sc}$.  Again $\frac{A_{\Delta_1}}{A_{\Delta}}$ and $\frac{A_{\Delta_2}}{A_{\Delta}}$ vanish at $J_K^0$ as shown in Fig.~\ref{fig:polaron}(c), confirming the absence of the polaron here. In the regime $J_K^0<J_K<J_K^c$, the amplitude for the polaron-polaron Cooper pair is the strongest. This again highlights the importance of the polaron for the PDW superconductor. 

 In summary, at $J_K^0$, the mixture of the polaron and the electron is the weakest because the inter-layer spin-spin correlation vanishes.  Moving away from $J_K^0$ to either the ferromagnetic and anti-ferromagnetic side, there is a hybridization between  electron and polaron. In the same time, the Fermi velocity decreases and the Luttinger parameter increases (see Fig.~\ref{fig:polaron}(d)), indicating effective attractive interactions. One simple explanation is that now the spin-spin exchange between the spin polarons has contributions from $J_c,J_s, J_{cs}$, which add up to induce a strong attraction between two nearby polarons.

\textbf{Bosonization analysis:}  The existence of the PDW phase can be understood from a bosonization analysis starting from the decoupled limit with $J_{cs}=J_K=0$ following Ref.~\onlinecite{berg2010pair,jaefari2012pair}.  At the decoupled limit we have one charge  and one spin mode from the C layer and an additional spin mode from the S layer. We can label the bosonization variables of the charge mode as $\theta_c, \phi_c$,  of the spin mode of the C layer as $\theta_s, \phi_s$ and of the spin mode in the S layer as  $\tilde \theta_s, \tilde \phi_s$. With the inclusion of $J_K, J_{cs}$, the two spin modes mix with each other and we can define new variable $\theta_{s;\pm}=\frac{1}{\sqrt{2}}(\theta_s\pm \tilde \theta_s)$ and $\phi_{s;\pm}=\frac{1}{\sqrt{2}}(\phi_s \pm \tilde \phi_s)$.  It can be shown that the most relevant inter-layer coupling term gives $-g \cos 2 \theta_{s;-} \cos 2 \phi_{s;+}$. In the supplementary\cite{SM} we show that this term is relevant when $J_K+2J_{cs}>0$ in the weak coupling limit. Therefore when $J_K+2 J_{cs}>0$, this term pins $\theta_{s;-}=0 \, (\pi)$ and $\phi_{s;+}=0\, (\pi)$ and the two spin modes are gapped. We are left with only the charge mode, leading to a spin gapped Luther-Emery liquid phase with algebraic superconductor (SC) and CDW order. 

However, correlation functions for simple $Q=0$ SC order defined in the C layer are exponentially decaying. We have spin-singlet SC order $\Delta_S\sim e^{i \sqrt{2}\theta_c} \cos \sqrt{2} \phi_s$ and spin-triplet SC order $\Delta_T\sim e^{i\sqrt{2} \theta_c}(\sin \sqrt{2}\theta_s, \cos \sqrt{2}\theta_s,\sin \sqrt{2}\phi_s)$. We note that $\phi_s=\frac{1}{\sqrt{2}}(\phi_{s;+}+\phi_{s;-})$ is always fluctuating because $\theta_{s;-}$ is pinned. Similarly $\theta_s$ is fluctuating because $\phi_{s;+}$ is ordered and all of these order parameters are gapped. To get algebraic decay, we need a composite order parameter by attaching an operator in the S layer.  First, in S layer we can define Neel order parameter  $(n_x,n_y,n_z)\sim (\sin \sqrt{2}\tilde \theta_s, \cos \sqrt{2}\tilde \theta_s,\sin \sqrt{2}\tilde \phi_s)$. Meanwhile, there is a VBS order parameter $\tilde V \sim \cos \sqrt{2}\tilde \phi_s$. The Neel and VBS order parameters carry momentum $\mathbf Q=\pi$. Now we can define a composite order parameter  $O_{\text{PDW}}\sim \Delta_S \tilde V \sim \vec{\Delta}_T \cdot \vec n  \sim e^{-i \sqrt{2} \theta_c}$ which carries momentum $\mathbf Q=\pi$ and is a spin-singlet. We have $  O_{\text{PDW}}(x) O_{\text{PDW}}(0) \sim \frac{1}{x^{\frac{1}{K_c}}} $.  Note that these composite order parameters $\Delta_S \tilde V $ and $\vec{\Delta}_T\cdot \vec n$, combined with a factor $(-1)^x$, are precisely $\Delta_1$ and $\Delta_2$ defined previously from electron-polaron Cooper pair and polaron-polaron Cooper pair.  

We need to emphasize that the above bosonization analysis starts from the decoupled limit and does not explain why $K_c$ can be larger than one as in our model. We believe that the formation of the spin polaron in the strong coupling regime  with $J_K$ deviating away from $J_K^0$ is crucial for an effective attractive interaction and is absent in the weak coupling analyses\cite{jaefari2012pair}.

\textbf{Type II t-J model:} Although a PDW phase at $J_K>0$ side can be explained by bosonization at least in the small $J_K$ limit, its existence at $J_K<0$ side is a surprise. In this subsection we show that a PDW phase could exist even at the $J_K \rightarrow -\infty$ limit. In the $J_K\rightarrow -\infty$ limit, we can obtain a type II t-J model which was recently proposed by us\cite{zhang2020type}. The model has two spin $1/2$ singlon and three $S=1$ doublon at each site. Here singlon is defined as singly occupied site and the doublon is defined as the doubly occupied site.  The model can be written as

 \begin{align}
 	H&=-t \sum_{ij;\sigma}c^\dagger_{i;\sigma} c_{j;\sigma}+J_s \sum_{\langle ij \rangle}\vec{s}_i \cdot \vec{s}_j\notag\\
 	&~~+J_d \sum_{\langle ij \rangle}\vec{S}_i \cdot \vec{S}_j+J_{sd}\sum_{\langle ij \rangle}(\vec{s}_i \cdot \vec{S}_j+\vec{S}_i\cdot \vec{s}_j)
 \end{align}
 where $\vec{s}$ is the spin operator of the singlon with $S=\frac{1}{2}$ and $\vec S$ is the spin operator of the doublon with $S=1$. The $c$ operator is projected to the restricted Hilbert space. We have $J_s=J_s$, $J_d=\frac{1}{2\sqrt{2}}(J_c+J_s+2J_{cs})$ and $J_{sd}=\frac{1}{2}(J_s+J_{cs})$. This model should be the effective t-J model from doping a $S=1$ Mott insulator with large Hund's coupling and crystal field splitting. On average, the $N_s$ atoms in the chain are each in the $d^{9-x}$ state, or equivalently, there are $(1-x) N_s$ number of $S=\frac{1}{2}$ sites in the $d^9$ configuration and $x N_s$ number of S=1 sites in the $d^8$ configuration.

\begin{figure}[ht]
\centering
\includegraphics[width=0.45\textwidth]{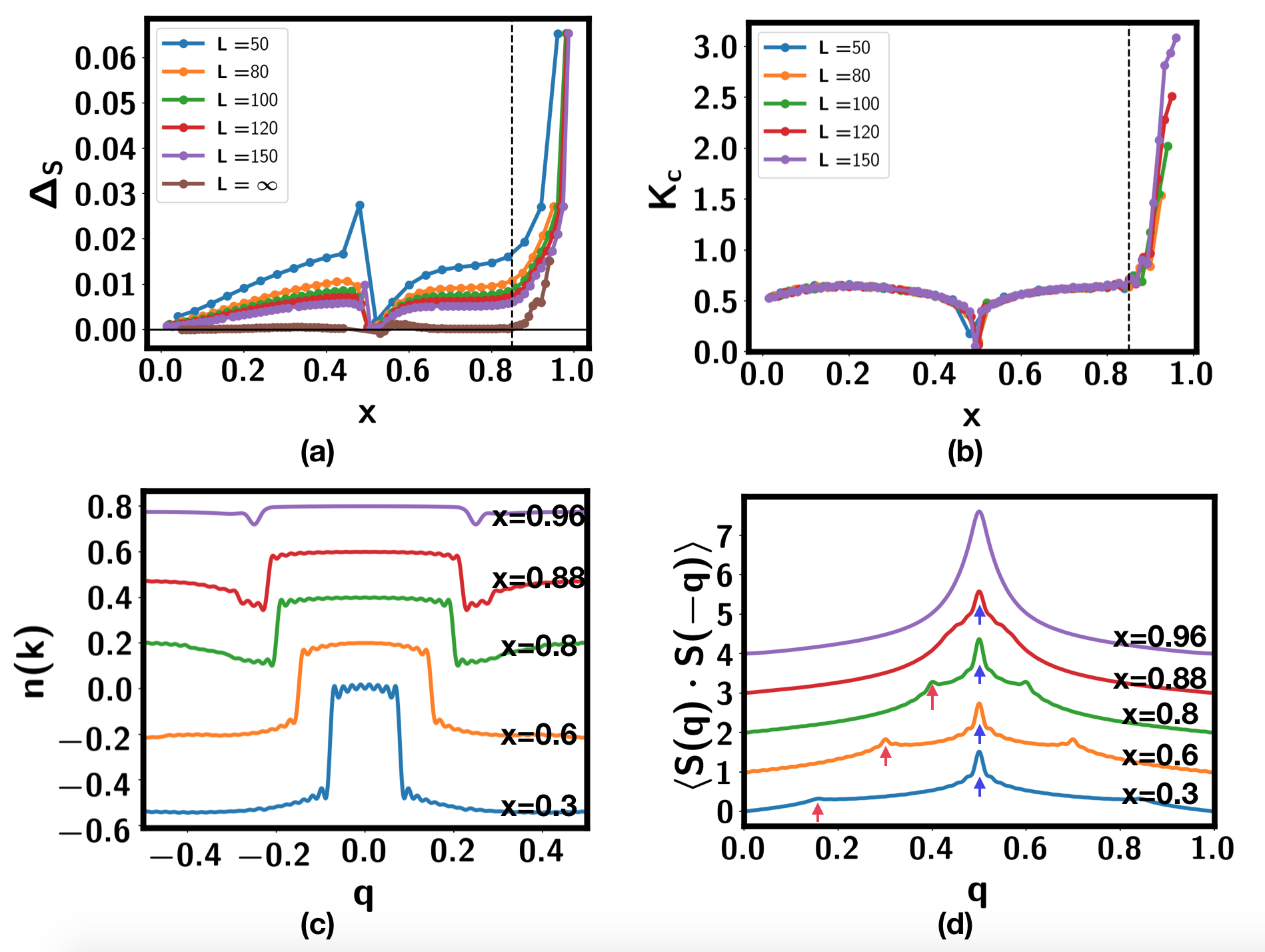}
\caption{Phase diagram of the type II t-J model with doping $x$, setting $t=1, J_s=J_d=0.5, J_{sd}=0.25$. The momentum in the plot is in units of $2\pi$. (a) Spin gap $\Delta_S$ with $x$. $\Delta_S$ at $L=\infty$ is extrapolated from that of the finite $L$. The dashed line is at $x_c=0.85$. (b) Luttinger parameter $K_c$ with $x$. (c) Momentum distribution function $n(k)=\langle c^\dagger_{\sigma}(k)c_\sigma(k)\rangle$. One can see a small pocket with $k_F=\frac{x}{4}2\pi$ when $x<x_c$.   (d) Spin-spin structure factor. Characteristic of LL*, there are two modes at momentum $2k_F$ and $\pi$ when $x<x_c$, denoted by red arrows and blue arrows respectively. }
\label{fig:type_II_main}
\end{figure}

While the rest of this paper has focused on doping close to unity, here a wider range of doping is displayed and the emergence of an unusual Luttinger liquid with small Luttinger volume is pointed out. In Fig.~\ref{fig:type_II_main} we show results for the type II t-J model.  We find a phase transition between a fractional Luttinger liquid (LL*) phase\cite{zhang2021fractional} ($x<x_c$) and the PDW superconductor ($x>x_c$) at around $x_c=0.85$. At $x=0.5$ a charge-density-wave (CDW) insulator is obtained.  The LL* phase has one spinful small Fermi surface with volume $2k_F=\frac{x}{2} 2\pi$ and an additional spin mode at momentum $\pi$(For details, see Ref.~\onlinecite{zhang2021fractional}).  The onset of the spin gap at $x_c$ is shown in Fig.~\ref{fig:type_II_main}(a). Meanwhile the Luttinger parameter $K_c$ becomes large when $x>x_c$ shown in Fig.~\ref{fig:type_II_main}(b), giving slow decay of pair-pair correlation function. The pair correlation function again reveals the oscillatory behavior of a PDW (see supplementary\cite{SM}) . Thus again we see that  a lightly doped Haldane chain in this large Hund's coupling limit, also reveals PDW superconductivity.

We can clearly see the expansion of the Fermi surface with $x$ when $x<x_c$ through the momentum distribution function $n(\mathbf k)=\langle c^\dagger_{\sigma}(\mathbf k)c_{\sigma}(\mathbf k)\rangle$ in Fig.~\ref{fig:type_II_main}(c).  The spin-spin structure factor $\langle \vec{S}(\mathbf q)\cdot \vec{S}(-\mathbf q) \rangle$ in Fig.~\ref{fig:type_II_main}(d) has peaks at both $2k_F=\frac{x}{2}$ and $\mathbf q=\pi$ in the LL* phase. When $x>x_c$, there is a spin gap and spin-spin structure factor only has a broad peak at $\mathbf q=\pi$, consistent with a spin gap. 

A key observation is that the LL* phase is qualitatively similar to the phase in the decoupled limit of the generalized Kondo model, although now we actually have $J_K=-\infty$. In the $J_K=-\infty$ limit, we can remove $J_K$ by dealing with the type II t-J model with a restricted Hilbert space.  It can be shown\cite{zhang2021fractional} that there are emergent orbitals which form effective $\tilde C$ layer and $\tilde S$ layer. In terms of the $\tilde C$ and $\tilde S$ layer, there is no Hund's coupling $-J_K$ anymore, as such a term does not exist in the type II t-J model. There can be effective anti-ferromagnetic spin-spin coupling between the $\tilde C$ and the $\tilde S$ layers coming from $J_s,J_d,J_{sd}>0$ terms in the type II t-J model. Such a coupling resembles an anti-ferromagnetic $J_{cs}$ coupling in terms of the new emergent orbitals and can drive the LL* phase into a PDW phase following the same bosonization analysis as in the weak coupling limit of the generalized Kondo model. In the supplementary\cite{SM} we also show that the central charge jumps from $c=3$ to $c=1$ across this transition, which is conjectured to be in the Kosterlitz-Thouless universality class.

\textbf{Conclusion} In summary, using a combination of numerical calculations and analytical arguments, we predict superconductivity, in fact a PDW i.e. a superconductor in which the Cooper pairs condensed with a non-zero center of mass momentum, on doping a spin-one Haldane chain or the rung singlet phase in a two leg ladder. Experimentally, the former may be realized by doping a spin-one chain formed by Ni$^{2+}$\cite{kojima1995musr} , while the latter can be realized by a two-leg ladder of fermionic atoms in an optical lattice, by preferentially doping one of the legs.  We show that the formation of a fermionic spin polaron is crucial for a robust PDW phase with slow decay of pairing correlation. Experimentally establishing PDW order is an interesting challenge, for example it may be revealed as a density wave when the sample is in contact with a conventional superconductor. A $\pi$ PDW on a ring with an odd number of sites, or more physically, a dislocation in a quasi 1-D crystal, should induce half-vortices due to the position-phase locking. Note, although we are here doping the Haldane chain, a paradigmatic example of an symmetry protected topological (SPT) phase, the edge modes have not played any role. It is left to future work if SPT physics has any role to play, given the presence of low energy charge excitations\cite{Rosch,Verresen}. The models studied here can be defined in any dimension, so the extension to two dimensions for instance should throw light on the intrinsic mechanism leading to PDWs  and the role of spin-polarons in mediating pairing in strongly correlated systems.

{\bf Acknowledgements:}  We would like to thank Annabelle Bohrdt, Ruben Verresen and Eslam Khalaf for discussions. Support from  Simons Collaboration on Ultra Quantum Matter, a grant from the Simons Foundation (651440, A.V.) and a Simons Investigator award are acknowledged.

\bibliographystyle{apsrev4-1}
\bibliography{pdw}

\begin{thebibliography}{30}%
\makeatletter
\providecommand \@ifxundefined [1]{%
 \@ifx{#1\undefined}
}%
\providecommand \@ifnum [1]{%
 \ifnum #1\expandafter \@firstoftwo
 \else \expandafter \@secondoftwo
 \fi
}%
\providecommand \@ifx [1]{%
 \ifx #1\expandafter \@firstoftwo
 \else \expandafter \@secondoftwo
 \fi
}%
\providecommand \natexlab [1]{#1}%
\providecommand \enquote  [1]{``#1''}%
\providecommand \bibnamefont  [1]{#1}%
\providecommand \bibfnamefont [1]{#1}%
\providecommand \citenamefont [1]{#1}%
\providecommand \href@noop [0]{\@secondoftwo}%
\providecommand \href [0]{\begingroup \@sanitize@url \@href}%
\providecommand \@href[1]{\@@startlink{#1}\@@href}%
\providecommand \@@href[1]{\endgroup#1\@@endlink}%
\providecommand \@sanitize@url [0]{\catcode `\\12\catcode `\$12\catcode
  `\&12\catcode `\#12\catcode `\^12\catcode `\_12\catcode `\%12\relax}%
\providecommand \@@startlink[1]{}%
\providecommand \@@endlink[0]{}%
\providecommand \url  [0]{\begingroup\@sanitize@url \@url }%
\providecommand \@url [1]{\endgroup\@href {#1}{\urlprefix }}%
\providecommand \urlprefix  [0]{URL }%
\providecommand \Eprint [0]{\href }%
\providecommand \doibase [0]{http://dx.doi.org/}%
\providecommand \selectlanguage [0]{\@gobble}%
\providecommand \bibinfo  [0]{\@secondoftwo}%
\providecommand \bibfield  [0]{\@secondoftwo}%
\providecommand \translation [1]{[#1]}%
\providecommand \BibitemOpen [0]{}%
\providecommand \bibitemStop [0]{}%
\providecommand \bibitemNoStop [0]{.\EOS\space}%
\providecommand \EOS [0]{\spacefactor3000\relax}%
\providecommand \BibitemShut  [1]{\csname bibitem#1\endcsname}%
\let\auto@bib@innerbib\@empty
\bibitem [{\citenamefont {Lee}\ \emph {et~al.}(2006)\citenamefont {Lee},
  \citenamefont {Nagaosa},\ and\ \citenamefont {Wen}}]{lee2006doping}%
  \BibitemOpen
  \bibfield  {author} {\bibinfo {author} {\bibfnamefont {P.~A.}\ \bibnamefont
  {Lee}}, \bibinfo {author} {\bibfnamefont {N.}~\bibnamefont {Nagaosa}}, \ and\
  \bibinfo {author} {\bibfnamefont {X.-G.}\ \bibnamefont {Wen}},\ }\href@noop
  {} {\bibfield  {journal} {\bibinfo  {journal} {Reviews of modern physics}\
  }\textbf {\bibinfo {volume} {78}},\ \bibinfo {pages} {17} (\bibinfo {year}
  {2006})}\BibitemShut {NoStop}%
\bibitem [{\citenamefont {Haldane}(1983)}]{haldane1983nonlinear}%
  \BibitemOpen
  \bibfield  {author} {\bibinfo {author} {\bibfnamefont {F.~D.~M.}\
  \bibnamefont {Haldane}},\ }\href@noop {} {\bibfield  {journal} {\bibinfo
  {journal} {Physical review letters}\ }\textbf {\bibinfo {volume} {50}},\
  \bibinfo {pages} {1153} (\bibinfo {year} {1983})}\BibitemShut {NoStop}%
\bibitem [{\citenamefont {Affleck}(1989)}]{affleck1989quantum}%
  \BibitemOpen
  \bibfield  {author} {\bibinfo {author} {\bibfnamefont {I.}~\bibnamefont
  {Affleck}},\ }\href@noop {} {\bibfield  {journal} {\bibinfo  {journal}
  {Journal of Physics: Condensed Matter}\ }\textbf {\bibinfo {volume} {1}},\
  \bibinfo {pages} {3047} (\bibinfo {year} {1989})}\BibitemShut {NoStop}%
\bibitem [{\citenamefont {Affleck}\ \emph {et~al.}(2004)\citenamefont
  {Affleck}, \citenamefont {Kennedy}, \citenamefont {Lieb},\ and\ \citenamefont
  {Tasaki}}]{affleck2004rigorous}%
  \BibitemOpen
  \bibfield  {author} {\bibinfo {author} {\bibfnamefont {I.}~\bibnamefont
  {Affleck}}, \bibinfo {author} {\bibfnamefont {T.}~\bibnamefont {Kennedy}},
  \bibinfo {author} {\bibfnamefont {E.~H.}\ \bibnamefont {Lieb}}, \ and\
  \bibinfo {author} {\bibfnamefont {H.}~\bibnamefont {Tasaki}},\ }in\
  \href@noop {} {\emph {\bibinfo {booktitle} {Condensed Matter Physics and
  Exactly Soluble Models}}}\ (\bibinfo  {publisher} {Springer},\ \bibinfo
  {year} {2004})\ pp.\ \bibinfo {pages} {249--252}\BibitemShut {NoStop}%
\bibitem [{\citenamefont {Gu}\ and\ \citenamefont {Wen}(2009)}]{gu2009tensor}%
  \BibitemOpen
  \bibfield  {author} {\bibinfo {author} {\bibfnamefont {Z.-C.}\ \bibnamefont
  {Gu}}\ and\ \bibinfo {author} {\bibfnamefont {X.-G.}\ \bibnamefont {Wen}},\
  }\href@noop {} {\bibfield  {journal} {\bibinfo  {journal} {Physical Review
  B}\ }\textbf {\bibinfo {volume} {80}},\ \bibinfo {pages} {155131} (\bibinfo
  {year} {2009})}\BibitemShut {NoStop}%
\bibitem [{\citenamefont {Pollmann}\ \emph {et~al.}(2010)\citenamefont
  {Pollmann}, \citenamefont {Turner}, \citenamefont {Berg},\ and\ \citenamefont
  {Oshikawa}}]{pollmann2010entanglement}%
  \BibitemOpen
  \bibfield  {author} {\bibinfo {author} {\bibfnamefont {F.}~\bibnamefont
  {Pollmann}}, \bibinfo {author} {\bibfnamefont {A.~M.}\ \bibnamefont
  {Turner}}, \bibinfo {author} {\bibfnamefont {E.}~\bibnamefont {Berg}}, \ and\
  \bibinfo {author} {\bibfnamefont {M.}~\bibnamefont {Oshikawa}},\ }\href@noop
  {} {\bibfield  {journal} {\bibinfo  {journal} {Physical review b}\ }\textbf
  {\bibinfo {volume} {81}},\ \bibinfo {pages} {064439} (\bibinfo {year}
  {2010})}\BibitemShut {NoStop}%
\bibitem [{\citenamefont {Fazekas}(1999)}]{fazekas1999lecture}%
  \BibitemOpen
  \bibfield  {author} {\bibinfo {author} {\bibfnamefont {P.}~\bibnamefont
  {Fazekas}},\ }\href@noop {} {\emph {\bibinfo {title} {Lecture notes on
  electron correlation and magnetism}}},\ Vol.~\bibinfo {volume} {5}\ (\bibinfo
   {publisher} {World scientific},\ \bibinfo {year} {1999})\BibitemShut
  {NoStop}%
\bibitem [{\citenamefont {Zhang}\ and\ \citenamefont
  {Vishwanath}(2020)}]{zhang2020type}%
  \BibitemOpen
  \bibfield  {author} {\bibinfo {author} {\bibfnamefont {Y.-H.}\ \bibnamefont
  {Zhang}}\ and\ \bibinfo {author} {\bibfnamefont {A.}~\bibnamefont
  {Vishwanath}},\ }\href@noop {} {\bibfield  {journal} {\bibinfo  {journal}
  {Physical Review Research}\ }\textbf {\bibinfo {volume} {2}},\ \bibinfo
  {pages} {023112} (\bibinfo {year} {2020})}\BibitemShut {NoStop}%
\bibitem [{\citenamefont {Zhang}\ and\ \citenamefont
  {Zhu}(2021)}]{zhang2021fractional}%
  \BibitemOpen
  \bibfield  {author} {\bibinfo {author} {\bibfnamefont {Y.-H.}\ \bibnamefont
  {Zhang}}\ and\ \bibinfo {author} {\bibfnamefont {Z.}~\bibnamefont {Zhu}},\
  }\href@noop {} {\bibfield  {journal} {\bibinfo  {journal} {Physical Review
  B}\ }\textbf {\bibinfo {volume} {103}},\ \bibinfo {pages} {115101} (\bibinfo
  {year} {2021})}\BibitemShut {NoStop}%
\bibitem [{\citenamefont {Gall}\ \emph {et~al.}(2021)\citenamefont {Gall},
  \citenamefont {Wurz}, \citenamefont {Samland}, \citenamefont {Chan},\ and\
  \citenamefont {K{\"o}hl}}]{gall2021competing}%
  \BibitemOpen
  \bibfield  {author} {\bibinfo {author} {\bibfnamefont {M.}~\bibnamefont
  {Gall}}, \bibinfo {author} {\bibfnamefont {N.}~\bibnamefont {Wurz}}, \bibinfo
  {author} {\bibfnamefont {J.}~\bibnamefont {Samland}}, \bibinfo {author}
  {\bibfnamefont {C.~F.}\ \bibnamefont {Chan}}, \ and\ \bibinfo {author}
  {\bibfnamefont {M.}~\bibnamefont {K{\"o}hl}},\ }\href@noop {} {\bibfield
  {journal} {\bibinfo  {journal} {Nature}\ }\textbf {\bibinfo {volume} {589}},\
  \bibinfo {pages} {40} (\bibinfo {year} {2021})}\BibitemShut {NoStop}%
\bibitem [{\citenamefont {Sompet}\ \emph {et~al.}(2021)\citenamefont {Sompet},
  \citenamefont {Hirthe}, \citenamefont {Bourgund}, \citenamefont {Chalopin},
  \citenamefont {Bibo}, \citenamefont {Koepsell}, \citenamefont {Bojovi{\'c}},
  \citenamefont {Verresen}, \citenamefont {Pollmann}, \citenamefont {Salomon}
  \emph {et~al.}}]{sompet2021realising}%
  \BibitemOpen
  \bibfield  {author} {\bibinfo {author} {\bibfnamefont {P.}~\bibnamefont
  {Sompet}}, \bibinfo {author} {\bibfnamefont {S.}~\bibnamefont {Hirthe}},
  \bibinfo {author} {\bibfnamefont {D.}~\bibnamefont {Bourgund}}, \bibinfo
  {author} {\bibfnamefont {T.}~\bibnamefont {Chalopin}}, \bibinfo {author}
  {\bibfnamefont {J.}~\bibnamefont {Bibo}}, \bibinfo {author} {\bibfnamefont
  {J.}~\bibnamefont {Koepsell}}, \bibinfo {author} {\bibfnamefont
  {P.}~\bibnamefont {Bojovi{\'c}}}, \bibinfo {author} {\bibfnamefont
  {R.}~\bibnamefont {Verresen}}, \bibinfo {author} {\bibfnamefont
  {F.}~\bibnamefont {Pollmann}}, \bibinfo {author} {\bibfnamefont
  {G.}~\bibnamefont {Salomon}},  \emph {et~al.},\ }\href@noop {} {\bibfield
  {journal} {\bibinfo  {journal} {arXiv preprint arXiv:2103.10421}\ } (\bibinfo
  {year} {2021})}\BibitemShut {NoStop}%
\bibitem [{\citenamefont {Agterberg}\ \emph {et~al.}(2020)\citenamefont
  {Agterberg}, \citenamefont {Davis}, \citenamefont {Edkins}, \citenamefont
  {Fradkin}, \citenamefont {Van~Harlingen}, \citenamefont {Kivelson},
  \citenamefont {Lee}, \citenamefont {Radzihovsky}, \citenamefont {Tranquada},\
  and\ \citenamefont {Wang}}]{agterberg2020physics}%
  \BibitemOpen
  \bibfield  {author} {\bibinfo {author} {\bibfnamefont {D.~F.}\ \bibnamefont
  {Agterberg}}, \bibinfo {author} {\bibfnamefont {J.~S.}\ \bibnamefont
  {Davis}}, \bibinfo {author} {\bibfnamefont {S.~D.}\ \bibnamefont {Edkins}},
  \bibinfo {author} {\bibfnamefont {E.}~\bibnamefont {Fradkin}}, \bibinfo
  {author} {\bibfnamefont {D.~J.}\ \bibnamefont {Van~Harlingen}}, \bibinfo
  {author} {\bibfnamefont {S.~A.}\ \bibnamefont {Kivelson}}, \bibinfo {author}
  {\bibfnamefont {P.~A.}\ \bibnamefont {Lee}}, \bibinfo {author} {\bibfnamefont
  {L.}~\bibnamefont {Radzihovsky}}, \bibinfo {author} {\bibfnamefont {J.~M.}\
  \bibnamefont {Tranquada}}, \ and\ \bibinfo {author} {\bibfnamefont
  {Y.}~\bibnamefont {Wang}},\ }\href@noop {} {\bibfield  {journal} {\bibinfo
  {journal} {Annual Review of Condensed Matter Physics}\ }\textbf {\bibinfo
  {volume} {11}},\ \bibinfo {pages} {231} (\bibinfo {year} {2020})}\BibitemShut
  {NoStop}%
\bibitem [{\citenamefont {Ning}\ \emph {et~al.}(2020)\citenamefont {Ning},
  \citenamefont {Liu},\ and\ \citenamefont {Jiang}}]{ning2020topological}%
  \BibitemOpen
  \bibfield  {author} {\bibinfo {author} {\bibfnamefont {S.-Q.}\ \bibnamefont
  {Ning}}, \bibinfo {author} {\bibfnamefont {Z.-X.}\ \bibnamefont {Liu}}, \
  and\ \bibinfo {author} {\bibfnamefont {H.-C.}\ \bibnamefont {Jiang}},\
  }\href@noop {} {\bibfield  {journal} {\bibinfo  {journal} {Physical Review
  Research}\ }\textbf {\bibinfo {volume} {2}},\ \bibinfo {pages} {023184}
  (\bibinfo {year} {2020})}\BibitemShut {NoStop}%
\bibitem [{\citenamefont {Keselman}\ \emph {et~al.}(2018)\citenamefont
  {Keselman}, \citenamefont {Berg},\ and\ \citenamefont
  {Azaria}}]{keselman2018one}%
  \BibitemOpen
  \bibfield  {author} {\bibinfo {author} {\bibfnamefont {A.}~\bibnamefont
  {Keselman}}, \bibinfo {author} {\bibfnamefont {E.}~\bibnamefont {Berg}}, \
  and\ \bibinfo {author} {\bibfnamefont {P.}~\bibnamefont {Azaria}},\
  }\href@noop {} {\bibfield  {journal} {\bibinfo  {journal} {Physical Review
  B}\ }\textbf {\bibinfo {volume} {98}},\ \bibinfo {pages} {214501} (\bibinfo
  {year} {2018})}\BibitemShut {NoStop}%
\bibitem [{\citenamefont {Zhu}\ \emph {et~al.}(2018)\citenamefont {Zhu},
  \citenamefont {Sheng},\ and\ \citenamefont {Weng}}]{zhu2018pairing}%
  \BibitemOpen
  \bibfield  {author} {\bibinfo {author} {\bibfnamefont {Z.}~\bibnamefont
  {Zhu}}, \bibinfo {author} {\bibfnamefont {D.}~\bibnamefont {Sheng}}, \ and\
  \bibinfo {author} {\bibfnamefont {Z.-Y.}\ \bibnamefont {Weng}},\ }\href@noop
  {} {\bibfield  {journal} {\bibinfo  {journal} {Physical Review B}\ }\textbf
  {\bibinfo {volume} {97}},\ \bibinfo {pages} {115144} (\bibinfo {year}
  {2018})}\BibitemShut {NoStop}%
\bibitem [{\citenamefont {Jiang}\ \emph {et~al.}(2018)\citenamefont {Jiang},
  \citenamefont {Li}, \citenamefont {Seidel},\ and\ \citenamefont
  {Lee}}]{jiang2018symmetry}%
  \BibitemOpen
  \bibfield  {author} {\bibinfo {author} {\bibfnamefont {H.-C.}\ \bibnamefont
  {Jiang}}, \bibinfo {author} {\bibfnamefont {Z.-X.}\ \bibnamefont {Li}},
  \bibinfo {author} {\bibfnamefont {A.}~\bibnamefont {Seidel}}, \ and\ \bibinfo
  {author} {\bibfnamefont {D.-H.}\ \bibnamefont {Lee}},\ }\href@noop {}
  {\bibfield  {journal} {\bibinfo  {journal} {Science bulletin}\ }\textbf
  {\bibinfo {volume} {63}},\ \bibinfo {pages} {753} (\bibinfo {year}
  {2018})}\BibitemShut {NoStop}%
\bibitem [{\citenamefont {Giamarchi}(2003)}]{giamarchi2003quantum}%
  \BibitemOpen
  \bibfield  {author} {\bibinfo {author} {\bibfnamefont {T.}~\bibnamefont
  {Giamarchi}},\ }\href@noop {} {\emph {\bibinfo {title} {Quantum physics in
  one dimension}}},\ Vol.\ \bibinfo {volume} {121}\ (\bibinfo  {publisher}
  {Clarendon press},\ \bibinfo {year} {2003})\BibitemShut {NoStop}%
\bibitem [{\citenamefont {Berg}\ \emph {et~al.}(2010)\citenamefont {Berg},
  \citenamefont {Fradkin},\ and\ \citenamefont {Kivelson}}]{berg2010pair}%
  \BibitemOpen
  \bibfield  {author} {\bibinfo {author} {\bibfnamefont {E.}~\bibnamefont
  {Berg}}, \bibinfo {author} {\bibfnamefont {E.}~\bibnamefont {Fradkin}}, \
  and\ \bibinfo {author} {\bibfnamefont {S.~A.}\ \bibnamefont {Kivelson}},\
  }\href@noop {} {\bibfield  {journal} {\bibinfo  {journal} {Physical review
  letters}\ }\textbf {\bibinfo {volume} {105}},\ \bibinfo {pages} {146403}
  (\bibinfo {year} {2010})}\BibitemShut {NoStop}%
\bibitem [{\citenamefont {Cho}\ \emph {et~al.}(2014)\citenamefont {Cho},
  \citenamefont {Soto-Garrido},\ and\ \citenamefont
  {Fradkin}}]{cho2014topological}%
  \BibitemOpen
  \bibfield  {author} {\bibinfo {author} {\bibfnamefont {G.~Y.}\ \bibnamefont
  {Cho}}, \bibinfo {author} {\bibfnamefont {R.}~\bibnamefont {Soto-Garrido}}, \
  and\ \bibinfo {author} {\bibfnamefont {E.}~\bibnamefont {Fradkin}},\
  }\href@noop {} {\bibfield  {journal} {\bibinfo  {journal} {Physical review
  letters}\ }\textbf {\bibinfo {volume} {113}},\ \bibinfo {pages} {256405}
  (\bibinfo {year} {2014})}\BibitemShut {NoStop}%
\bibitem [{\citenamefont {May-Mann}\ \emph {et~al.}(2020)\citenamefont
  {May-Mann}, \citenamefont {Levy}, \citenamefont {Soto-Garrido}, \citenamefont
  {Cho}, \citenamefont {Clark},\ and\ \citenamefont
  {Fradkin}}]{may2020topology}%
  \BibitemOpen
  \bibfield  {author} {\bibinfo {author} {\bibfnamefont {J.}~\bibnamefont
  {May-Mann}}, \bibinfo {author} {\bibfnamefont {R.}~\bibnamefont {Levy}},
  \bibinfo {author} {\bibfnamefont {R.}~\bibnamefont {Soto-Garrido}}, \bibinfo
  {author} {\bibfnamefont {G.~Y.}\ \bibnamefont {Cho}}, \bibinfo {author}
  {\bibfnamefont {B.~K.}\ \bibnamefont {Clark}}, \ and\ \bibinfo {author}
  {\bibfnamefont {E.}~\bibnamefont {Fradkin}},\ }\href@noop {} {\bibfield
  {journal} {\bibinfo  {journal} {Physical Review B}\ }\textbf {\bibinfo
  {volume} {101}},\ \bibinfo {pages} {165133} (\bibinfo {year}
  {2020})}\BibitemShut {NoStop}%
\bibitem [{\citenamefont {Bohrdt}\ \emph
  {et~al.}(2021{\natexlab{a}})\citenamefont {Bohrdt}, \citenamefont {Homeier},
  \citenamefont {Bloch}, \citenamefont {Demler},\ and\ \citenamefont
  {Grusdt}}]{bohrdt2021strong}%
  \BibitemOpen
  \bibfield  {author} {\bibinfo {author} {\bibfnamefont {A.}~\bibnamefont
  {Bohrdt}}, \bibinfo {author} {\bibfnamefont {L.}~\bibnamefont {Homeier}},
  \bibinfo {author} {\bibfnamefont {I.}~\bibnamefont {Bloch}}, \bibinfo
  {author} {\bibfnamefont {E.}~\bibnamefont {Demler}}, \ and\ \bibinfo {author}
  {\bibfnamefont {F.}~\bibnamefont {Grusdt}},\ }\href@noop {} {\bibfield
  {journal} {\bibinfo  {journal} {arXiv preprint arXiv:2108.04118}\ } (\bibinfo
  {year} {2021}{\natexlab{a}})}\BibitemShut {NoStop}%
\bibitem [{\citenamefont {Bohrdt}\ \emph
  {et~al.}(2021{\natexlab{b}})\citenamefont {Bohrdt}, \citenamefont {Homeier},
  \citenamefont {Reinmoser}, \citenamefont {Demler},\ and\ \citenamefont
  {Grusdt}}]{bohrdt2021exploration}%
  \BibitemOpen
  \bibfield  {author} {\bibinfo {author} {\bibfnamefont {A.}~\bibnamefont
  {Bohrdt}}, \bibinfo {author} {\bibfnamefont {L.}~\bibnamefont {Homeier}},
  \bibinfo {author} {\bibfnamefont {C.}~\bibnamefont {Reinmoser}}, \bibinfo
  {author} {\bibfnamefont {E.}~\bibnamefont {Demler}}, \ and\ \bibinfo {author}
  {\bibfnamefont {F.}~\bibnamefont {Grusdt}},\ }\href@noop {} {\bibfield
  {journal} {\bibinfo  {journal} {Annals of Physics}\ }\textbf {\bibinfo
  {volume} {435}},\ \bibinfo {pages} {168651} (\bibinfo {year}
  {2021}{\natexlab{b}})}\BibitemShut {NoStop}%
\bibitem [{\citenamefont {Kagan}\ \emph {et~al.}(1999)\citenamefont {Kagan},
  \citenamefont {Khomskii},\ and\ \citenamefont {Mostovoy}}]{kagan1999double}%
  \BibitemOpen
  \bibfield  {author} {\bibinfo {author} {\bibfnamefont {M.~Y.}\ \bibnamefont
  {Kagan}}, \bibinfo {author} {\bibfnamefont {D.}~\bibnamefont {Khomskii}}, \
  and\ \bibinfo {author} {\bibfnamefont {M.}~\bibnamefont {Mostovoy}},\
  }\href@noop {} {\bibfield  {journal} {\bibinfo  {journal} {The European
  Physical Journal B-Condensed Matter and Complex Systems}\ }\textbf {\bibinfo
  {volume} {12}},\ \bibinfo {pages} {217} (\bibinfo {year} {1999})}\BibitemShut
  {NoStop}%
\bibitem [{SM()}]{SM}%
  \BibitemOpen
  \href@noop {} {\bibinfo  {journal} {Supplementary material}\ }\BibitemShut
  {NoStop}%
\bibitem [{\citenamefont {Jaefari}\ and\ \citenamefont
  {Fradkin}(2012)}]{jaefari2012pair}%
  \BibitemOpen
\bibfield  {journal} {  }\bibfield  {author} {\bibinfo {author} {\bibfnamefont
  {A.}~\bibnamefont {Jaefari}}\ and\ \bibinfo {author} {\bibfnamefont
  {E.}~\bibnamefont {Fradkin}},\ }\href@noop {} {\bibfield  {journal} {\bibinfo
   {journal} {Physical Review B}\ }\textbf {\bibinfo {volume} {85}},\ \bibinfo
  {pages} {035104} (\bibinfo {year} {2012})}\BibitemShut {NoStop}%
\bibitem [{\citenamefont {Kojima}\ \emph {et~al.}(1995)\citenamefont {Kojima},
  \citenamefont {Keren}, \citenamefont {Le}, \citenamefont {Luke},
  \citenamefont {Wu}, \citenamefont {Uemura}, \citenamefont {Kiyono},
  \citenamefont {Miyasaka}, \citenamefont {Takagi},\ and\ \citenamefont
  {Uchida}}]{kojima1995musr}%
  \BibitemOpen
  \bibfield  {author} {\bibinfo {author} {\bibfnamefont {K.}~\bibnamefont
  {Kojima}}, \bibinfo {author} {\bibfnamefont {A.}~\bibnamefont {Keren}},
  \bibinfo {author} {\bibfnamefont {L.}~\bibnamefont {Le}}, \bibinfo {author}
  {\bibfnamefont {G.}~\bibnamefont {Luke}}, \bibinfo {author} {\bibfnamefont
  {W.}~\bibnamefont {Wu}}, \bibinfo {author} {\bibfnamefont {Y.}~\bibnamefont
  {Uemura}}, \bibinfo {author} {\bibfnamefont {K.}~\bibnamefont {Kiyono}},
  \bibinfo {author} {\bibfnamefont {S.}~\bibnamefont {Miyasaka}}, \bibinfo
  {author} {\bibfnamefont {H.}~\bibnamefont {Takagi}}, \ and\ \bibinfo {author}
  {\bibfnamefont {S.}~\bibnamefont {Uchida}},\ }\href@noop {} {\bibfield
  {journal} {\bibinfo  {journal} {Journal of magnetism and magnetic materials}\
  }\textbf {\bibinfo {volume} {140}},\ \bibinfo {pages} {1657} (\bibinfo {year}
  {1995})}\BibitemShut {NoStop}%
\bibitem [{\citenamefont {Anfuso}\ and\ \citenamefont {Rosch}(2007)}]{Rosch}%
  \BibitemOpen
  \bibfield  {author} {\bibinfo {author} {\bibfnamefont {F.}~\bibnamefont
  {Anfuso}}\ and\ \bibinfo {author} {\bibfnamefont {A.}~\bibnamefont {Rosch}},\
  }\href {\doibase 10.1103/physrevb.75.144420} {\bibfield  {journal} {\bibinfo
  {journal} {Physical Review B}\ }\textbf {\bibinfo {volume} {75}} (\bibinfo
  {year} {2007}),\ 10.1103/physrevb.75.144420}\BibitemShut {NoStop}%
\bibitem [{\citenamefont {Verresen}\ \emph {et~al.}(2021)\citenamefont
  {Verresen}, \citenamefont {Bibo},\ and\ \citenamefont {Pollmann}}]{Verresen}%
  \BibitemOpen
  \bibfield  {author} {\bibinfo {author} {\bibfnamefont {R.}~\bibnamefont
  {Verresen}}, \bibinfo {author} {\bibfnamefont {J.}~\bibnamefont {Bibo}}, \
  and\ \bibinfo {author} {\bibfnamefont {F.}~\bibnamefont {Pollmann}},\
  }\href@noop {} {\enquote {\bibinfo {title} {Quotient symmetry protected
  topological phenomena},}\ } (\bibinfo {year} {2021}),\ \Eprint
  {http://arxiv.org/abs/2102.08967} {arXiv:2102.08967 [cond-mat.str-el]}
  \BibitemShut {NoStop}%
\bibitem [{\citenamefont {Ogata}\ \emph {et~al.}(1991)\citenamefont {Ogata},
  \citenamefont {Luchini}, \citenamefont {Sorella},\ and\ \citenamefont
  {Assaad}}]{ogata1991phase}%
  \BibitemOpen
  \bibfield  {author} {\bibinfo {author} {\bibfnamefont {M.}~\bibnamefont
  {Ogata}}, \bibinfo {author} {\bibfnamefont {M.}~\bibnamefont {Luchini}},
  \bibinfo {author} {\bibfnamefont {S.}~\bibnamefont {Sorella}}, \ and\
  \bibinfo {author} {\bibfnamefont {F.}~\bibnamefont {Assaad}},\ }\href@noop {}
  {\bibfield  {journal} {\bibinfo  {journal} {Physical review letters}\
  }\textbf {\bibinfo {volume} {66}},\ \bibinfo {pages} {2388} (\bibinfo {year}
  {1991})}\BibitemShut {NoStop}%
\bibitem [{Note1()}]{Note1}%
  \BibitemOpen
  \bibinfo {note} {Note that $\eta _a \eta _b=\pm i$ because $(\eta _a \eta
  _b)^2=1$. In the end we only care about the correlation function of the spin
  operators, where a pair of $\eta _a \eta _b$ appear. Thus whether $\eta _a
  \eta _b=i$ or $\eta _a\eta _b=-i$ does not matter. Here we make one choice
  just for simplicity.}\BibitemShut {Stop}%
\end{thebibliography}%

\appendix
\onecolumngrid

\section{Realization of the generalized Kondo model and type II t-J model in solid state system}

\subsection{Two-orbital Hubbard model}

 We want to model a transition metal oxide with 3d electrons in atomic configuration $d^{9-x}$. We consider a model with two orbitals (for example, the two $e_g$ orbitals). We will use the hole picture for simplicity. A general lattice Hamiltonian is

\begin{align}
	H&=H_K+\frac{U_1}{2}\sum_i n_{1;i}(n_{1;i}-1)+\frac{U_2}{2}\sum_i n_{2;i}(n_{2;i}-1)+U'\sum_i n_{1;i}n_{2;i}-2J_H \sum_i (\mathbf{S}_{1;i}\cdot \mathbf{S}_{2;i}+\frac{1}{4}n_{i;1}n_{i;2})
\label{eq:spin_orbital_model}
\end{align}
where $n_{a;i}$ is the density of the orbital $a$ at the site $i$. $a=1,2$ denotes the $d_{z^z}$ orbital and the $d_{x^2-y^2}$ orbital respectively. In certain material, the energy of another orbital such as $d_{xy}$ is lower than that of the $d_{z^2}$ orbital. In this case we just use $d_1$ to represent the $d_{xy}$ orbital.   $U_1$, $U_2$ are intra-orbital Hubbard interaction. $U'$ is the inter-orbital interaction. $J_H$ is the inter-orbital Hund's coupling.  We assume that $U_1=U_2=U$ and $U-U'=2J_H$.

The kinetic energy is
\begin{align}
	H_K&=\sum_i \Delta n_{1;i}- \sum_{ ij } t_{1;ij} d^\dagger_{1;i}d_{1;j}- \sum_{ ij } t_{2;ij} d^\dagger_{2;i}d_{2;j}-\sum_{\langle ij \rangle} t_{12;ij}(d^\dagger_{1;i}d_{2;j}+h.c.)
	\label{eq:full_Hamiltonian}
\end{align}
where $\Delta$ is the crystal field splitting between the two  orbitals.

We consider the limit that $\Delta \geq t$ but $\Delta< J_H, U', U$. We label the total density per site as $n=1+x$. At $x=0$, we have one particle per site. Because of a finite $\Delta$, there are only two possible states: $d_{2;\uparrow}^\dagger \ket{0}$ and $d_{2;\downarrow}^\dagger \ket{0}$, forming a $S=1/2$ local moment. When $x>0$, there are $x N_s$ number of sites with $n=2$, where $N_s$ is the total number of sites.  If $U-U'+J_H=3J_H>\Delta$, the doped electron is favored to enter the $d_1$ orbital to reduce repulsion and gain from Hund's coupling. In the end, $d_2$ orbital is always frozen and remains as spin $1/2$ local moment.  We then reach a kondo like model with one orbital in a Mott insulating phase, which then reduces to the type II t-J model in the large $J_H$ limit.   If on the other hand $\Delta>3J_H$, the doped electron is favored to enter the $d_2$ orbital and forms a spin-singlet doublon, leading to the conventional type I t-J model previously studied.  In this paper we will restrict to  the regime that $\Delta<3J_H$.

\subsection{Generalized Kondo model}

\begin{figure}[ht]
\centering
\includegraphics[width=0.8\textwidth]{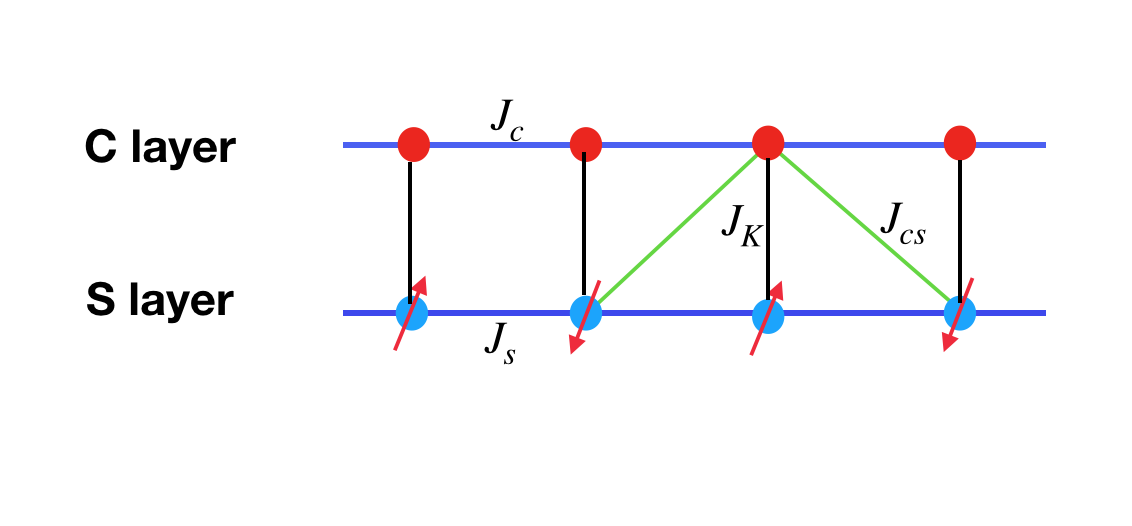}
\caption{Illustration of the generalized Kondo model defined in Eq.~\ref{eq:kondo_t_J}. The C layer corresponds to the $d_1$ orbital, while the S layer corresponds to the $d_2$ orbital which is Mott localized. $J_c,J_s, J_{cs}$ are anti-ferromagnetic super-exchange terms. $J_K$ is the on-site Kondo coupling. In transition metal oxides such as the nickelates, $J_K=-J_H$ is ferromagnetic and originates from the  inter-orbital Hund's coupling. On the other hand, we can also realize the model in a two leg optical lattice in cold atom system, there $J_K$ is also from the super-exchange process and is generically anti-ferromagnetic.}
\label{fig:kdon_hamiltonian_appendix}
\end{figure}

If $\Delta<3J_H$, the doped hole starting from the $d^9$ state will only enter the $d_1$ orbital while the $d_2$ orbital is always singly occupied. In another word, the $d_2$ orbital is orbitally selective Mott localized. Then we are left with a kondo like model where a conventional $t-J$ model coupled to spin $1/2$ local moment. We assume $U,U'>>t$, so the itinerant electron in the $d_1$ orbital itself is strongly correlated and is described by a $t-J$ model, which then couples to the local spin $1/2$ moment from the $d_2$ orbital. We then define $c_{i;\sigma}= P d_{i;1\sigma} P$, where $P$ is the projection operator to remove the double occupancy. Then we reach a Kondo-like model:

\begin{align}
H&=-t \sum_{ij}c^\dagger_i c_j+J_c \sum_{\langle ij \rangle}\vec{S}_{i;c} \cdot \vec{S}_{j;c}+J_s \sum_{\langle ij \rangle}\vec{S}_{i} \cdot \vec{S}_{j}+J_K \sum_i \vec{S}_{i;c}\cdot \vec{S}_i+J_{cs} \sum_{ij}\vec{S}_{i;c}\cdot \vec{S}_j
\label{eq:kondo_t_J_appendix}
\end{align}
where $J_K=-J_H$ is the ferromagnetic on-site Hund's coupling.  The first two terms are the conventional spin $1/2$ t-J model.  The third term is the Heisenberg model of the spin $1/2$ localized spin.  The last line is the spin-spin coupling between the itinerant electron and the localized spin. $J_c$ is the super-exchange term between the itinerant electron. $J_s$ is the super-exchange between the localized spins.  $J_{cs}$ is the super-exchange between the itinerant electron and the localized spin.  The model is illustrated in Fig.~\ref{fig:kdon_hamiltonian_appendix}. In the above we have ignored the possible $n_i n_j$ terms.

In the above, we have:

\begin{align}
J_c&=4 \frac{t_1^2}{U} \notag\\
J_s&=4 \frac{t_2^2}{U} \notag\\
J_{cs}&=2\frac{t_{12}^2}{U-U'-\Delta}+2\frac{t_{12}^2}{U+U'+\Delta}\notag\\
J_K&=-J_H
\end{align}
which is derived by assuming $J_H$ is small compared to $U', U$.

 In principle we should also include some three-site correlated hopping processes from $t^2/U$.  We will ignore them following the same procedure in the usual t-J model. They are listed below as:

 \begin{align}
H'=&-\frac{J_c}{4} \sum_i \left(c^\dagger_{i+1;\sigma}c_{i-1;\sigma} n_{i;\bar \sigma}-c^\dagger_{i+1;\uparrow}c_{i;\downarrow}S_{i;c}^--c^\dagger_{i+1;\downarrow}c_{i;\uparrow}S_{i;c}^+\right)+h.c.\notag\\
&-\sum_{ij}\left((\frac{2t_{12}^2}{U-U'-\Delta}-\frac{ 2t_{12}^2}{U-\Delta})-(2\frac{t_{12}^2}{U+\Delta}-\frac{2 t_{12}^2}{U+\Delta+U'})\right)n_j \vec{S}_{i;c}\cdot \vec{S_{j}}\notag\\
&+\sum_{\langle ij \rangle }\left(-\frac{t_1^2}{U}+(\frac{t_{12}^2}{U-U'-\Delta}-\frac{ t_{12}^2}{U-\Delta})-(\frac{t_{12}^2}{U+\Delta}-\frac{ t_{12}^2}{U+\Delta+U'})\right)n_i n_j \notag\\
&+\frac{1}{2} \sum_i (c^\dagger_{i+1;\sigma} c_{i-1;\sigma}) \left((\frac{t_{12}^2}{U'+\Delta}-\frac{t_{12}^2}{U-U'-\Delta})+(\frac{t_{12}^2}{\Delta}-\frac{t_{12}^2}{U'+\Delta}+\frac{t_{12}^2}{U-U'-\Delta}-\frac{t_{12}^2}{U-\Delta})n_i\right)\notag\\
&+\sum_i (c^\dagger_{i+1;\sigma}\vec{\sigma}_{\sigma\sigma'}c_{i-1;\sigma'})\cdot \vec{S}_i\left((\frac{t_{12}^2}{U'+\Delta}+\frac{t_{12}^2}{U-U'-\Delta})+(\frac{t_{12}^2}{\Delta}-\frac{t_{12}^2}{U'+\Delta}-\frac{t_{12}^2}{U-U'-\Delta}+\frac{t_{12}^2}{U-\Delta})n_i\right)
\end{align}

In the above we have hidden the projection operator to impose the constraint that there is no double occupancy in the C layer.

\subsection{Type II t-J model}

The type I and type-II t-J model can be reached by taking $J_K\rightarrow +\infty$ limit and $J_K \rightarrow -\infty$ limit respectively.  Let us take the $J_K \rightarrow -\infty$ limit, then we need to remove the inter-orbital spin singlet from the Hilbert space and get the type II t-J model\cite{zhang2020type,zhang2021fractional} with two spin $1/2$ singlon and three $S=1$ doublon at each site. Here singlon is defined as singly occupied site and the doublon is defined as the doubly occupied site.  The model can be written as

 \begin{align}
 	H&=-t \sum_{ij}c^\dagger_i c_j+J_s \sum_{\langle ij \rangle}\vec{s}_i \cdot \vec{s}_j\notag+J_d \sum_{\langle ij \rangle}\vec{S}_i \cdot \vec{S}_j+J_{sd}\sum_{\langle ij \rangle}(\vec{s}_i \cdot \vec{S}_j+\vec{S}_i\cdot \vec{s}_j)
 \end{align}
 where $\vec{s}$ is the spin operator of the singlon with $S=\frac{1}{2}$ and $\vec S$ is the spin operator of the doublon with $S=1$. We have

 \begin{align}
J_s&=J_s \notag\\
J_d&=\frac{1}{2\sqrt{2}}(J_c+J_s+2J_{cs})\notag\\
J_{sd}&=\frac{1}{2}(J_s+J_{cs})\notag\\
 \end{align}

\section{Realization of the generalized Kondo model in bilayer optical lattice}

Here we show that the generalized Kondo model with $J_K>0$ can also be naturally realized in a bilayer optical lattice system.  We consider a bilayer optical lattice described by a Hubbard model:

\begin{align}
H&=\Delta\sum_i n_{i;1}-t\sum_{a=1,2}\sum_{\sigma=\uparrow,\downarrow}\sum_{ij} c^\dagger_{i;a \sigma}c_{j;a \sigma}-t_{12}\sum_{a=1,2}\sum_{\sigma=\uparrow,\downarrow}\sum_{ij}(c^\dagger_{i;1\sigma}c_{j;2\sigma}+c^\dagger_{i;2\sigma}c_{j;1\sigma})-t_{\perp}\sum_{a,\sigma}\sum_{i}(c^\dagger_{i;1\sigma}c_{i;2\sigma}+h.c.)\notag\\
&-\mu \sum_{a=1,2} \sum_i n_{i;a}+\frac{U}{2}\sum_{a}\sum_i n_{i;a}(n_{i;a}-1)+U'\sum_i n_{i;1}n_{i;2}
\end{align}
where $n_{i;a}=\sum_{\sigma}c^\dagger_{i;a\sigma}c_{i;a\sigma}$ is the density at site $i$ for layer $a=1,2$. $n_i=n_{i;1}+n_{i;2}$ is the total density at site $i$.  We also define the average density $n=\frac{1}{N_s}\sum_i n_i$, where $N_s$ is the total number of sites in the system.

The model resembles the two-orbital Hubbard model in Eq.~\ref{eq:full_Hamiltonian}, but now we have $J_H=0$. Here $a=1,2$ labels the two layers and $t_\perp$ is the inter-layer vertical tunneling. A non-zero $\Delta>0$ is caused by a displacement field, or a potential difference between the two layers.  We will stay in the limit $U>>t$ and $U>>U'$. We assume $t_\perp,t<\Delta<U-U'$. At density $n=1$, we have a Mott insulator with one particle at the layer $2$. Then at density $n=1+x$ with $x \in (0,1)$, the doped additional particle enters the layer $1$ to reduce the on-site Hubbard U. In this case the layer $2$ is always Mott localized and provides a spin $1/2$ moment. The itinerant electron in the layer $1$ is described by a $t-J$ model which then couples to the local moment of the layer $2$ through a Kondo coupling.  This is exactly the generalized Kondo model defined in Eq.~\ref{eq:kondo_t_J_appendix} with the parameter:

\begin{align}
J_c&=4\frac{t^2}{U} \notag\\
J_s&=4\frac{t^2}{U}\notag\\
J_{cs}&=2\frac{t_{12}^2}{U-U'-\Delta}+2\frac{t_{12}^2}{U+U'+\Delta}\notag\\
J_K&=2\frac{t_{\perp}^2}{U-U'-\Delta}+2\frac{t_{\perp}^2}{U-U'+\Delta}
\end{align}

Note that we always have $J_K>0$ because we need $\Delta<U-U'$ to make the doped particles stay in the layer $1$.   If we increase either $\Delta$ or $t_{\perp}$, we should reach a Fermi liquid phase with large Fermi surface. We will mainly focus on the regime where $\Delta$ is not large enough to destroy the layer selective mott transition.. In this regime we can focus on the generalized Kondo model in Eq.~\ref{eq:kondo_t_J_appendix} with $J_K$ controlled by $t_{\perp}$.  Note in the above analysis $U'$ is not necessary and we can set $U'=0$.  $t_{12}$ is needed to generate a finite $J_{cs}$, which is not necessary, but can enhance the PDW phase.

\section{Extraction of Luttinger parameter and compressibility in DMRG \label{appendix:extraction_luttinger_parameter}}

We use the following formula to extract  the Luttinger parameter $K_c$ for the charge mode:

\begin{equation}
  N(\mathbf q)=\langle n(\mathbf q) n(-\mathbf q) \rangle= N(\mathbf q=0)+\frac{K_{\rho}}{\pi} q 
\end{equation}
when $q \rightarrow 0$.

Here, we do the Fourier transformation of

\begin{equation}
  \langle n(\mathbf x) n(0) \rangle =-\frac{K_\rho}{\pi^2}\frac{1}{x^2}
\end{equation}

to get

\begin{equation}
  N(\mathbf q)= \sum_{\mathbf x=0}^{L}\langle n(\mathbf x) n(0) \rangle
\end{equation}

We also try to extract the charge compressibility $\kappa_c$.  It is known that

\begin{equation}
  \frac{1}{\kappa_c}=\frac{\partial \mu}{\partial n}=\frac{\partial^2 E}{\partial n^2}=\frac{1}{L}\frac{E(n+\delta n)+E(n-\delta n)-2 E(n)}{\delta n^2}
\end{equation}

Similarly, there is a spin compressibility $\kappa_s$ extracted from:

\begin{equation}
  \frac{1}{\kappa_s}=\frac{\partial^2 E}{\partial S_z^2}=\frac{1}{L} \frac{E(S_z=\frac{1}{2})+E(S_z=\frac{1}{2})-2 E(S_z=0)}{ (\frac{1}{2}/L)^2}=8 \Delta_S L
\end{equation}

In Luttinger liquid theory, it is known that:
\begin{equation}
  \kappa_c=\frac{2 K_c}{\pi \upsilon_c}
\end{equation}

\begin{figure}[H]
\centering
\includegraphics[width=0.85\textwidth]{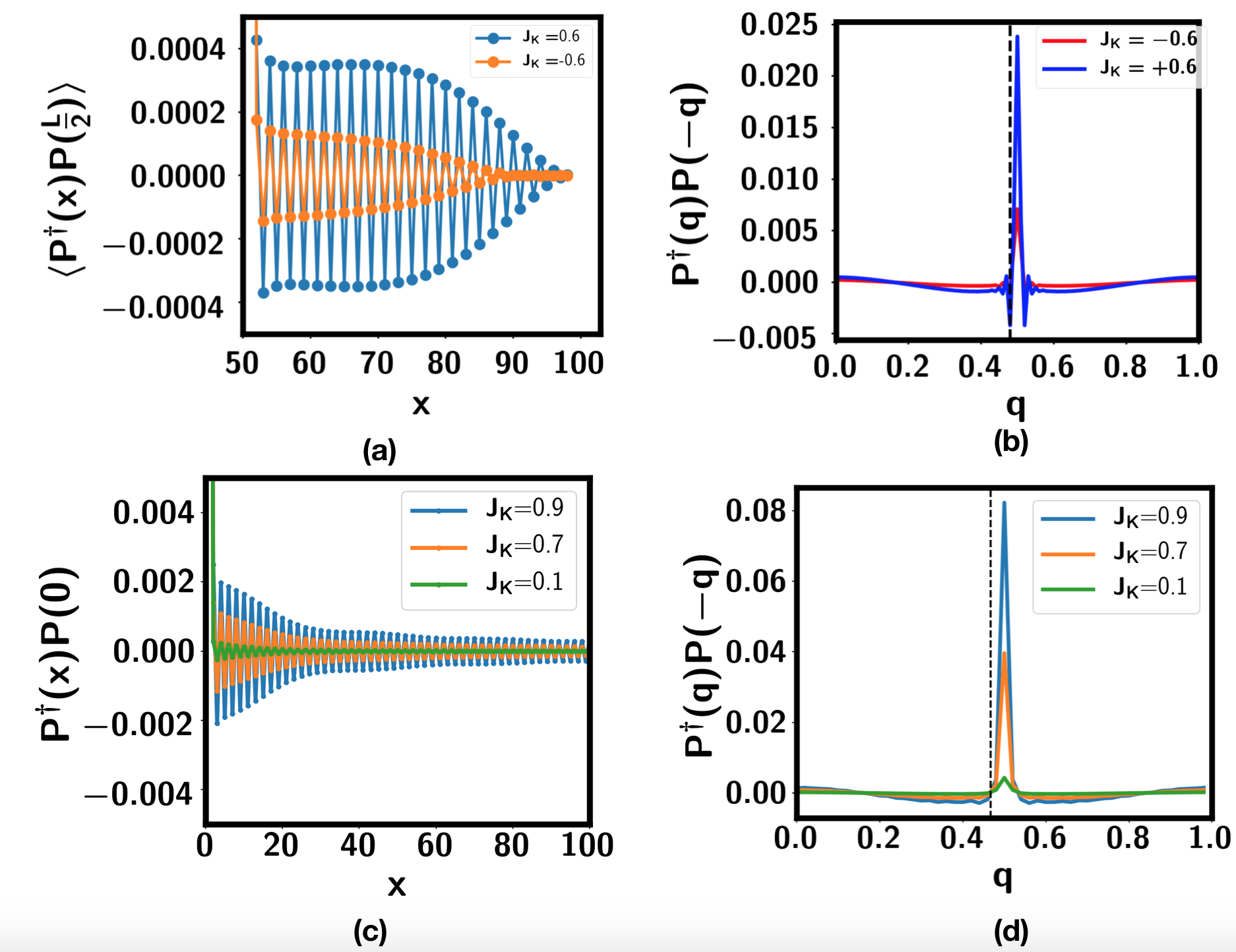}
\caption{We show pair-pair correlation function for $J_{cs}=0.25$ from finite DMRG and infinite DMRG. (a)(b) Pair-pair correlation function in real and momentum space from finite DMRG. $x=0.96$ with system size $L_x=100$.  (c)(d) Pair-pair correlation function in real and momentum space from infinite DMRG with unit cell size $L=30$ and $x=\frac{28}{30}$. The vertical dashed lines are at $q=2k_F=\frac{x}{2} \times 2\pi$. In the Fourier transformation of $P^\dagger(x)P(0)$, we ignored the short distance contribution with $|x|=0,1$. }
\label{fig:pair_pair_appendix}
\end{figure}

\section{More results on the generalized Kondo model}

\subsection{Pairing correlation in real space and momentum space}

In Fig.~\ref{fig:pair_pair_appendix} we show the comparison of pair-pair correlation function obtained from finite and infinite DMRG.  In finite DMRG, we see that the pair-pair correlation function has a sharp drop at the boundary of the system.  This turns out to enhance a peak in the fourier transform at momentum $q=2k_F=\frac{x}{2} \times 2\pi=\frac{48}{100} \times 2\pi$. We note that the $2k_F$ part in the pair-pair correlation function should have a large decay exponent $K_c+\frac{1}{K_c}>2$ and should be much smaller than the peak at $q=\pi$. Indeed, in infinite DMRG, we find the feature at $q=2k_F$ is significantly weaker because there is no boundary effect.

\begin{figure}[ht]
\centering
\includegraphics[width=0.5\textwidth]{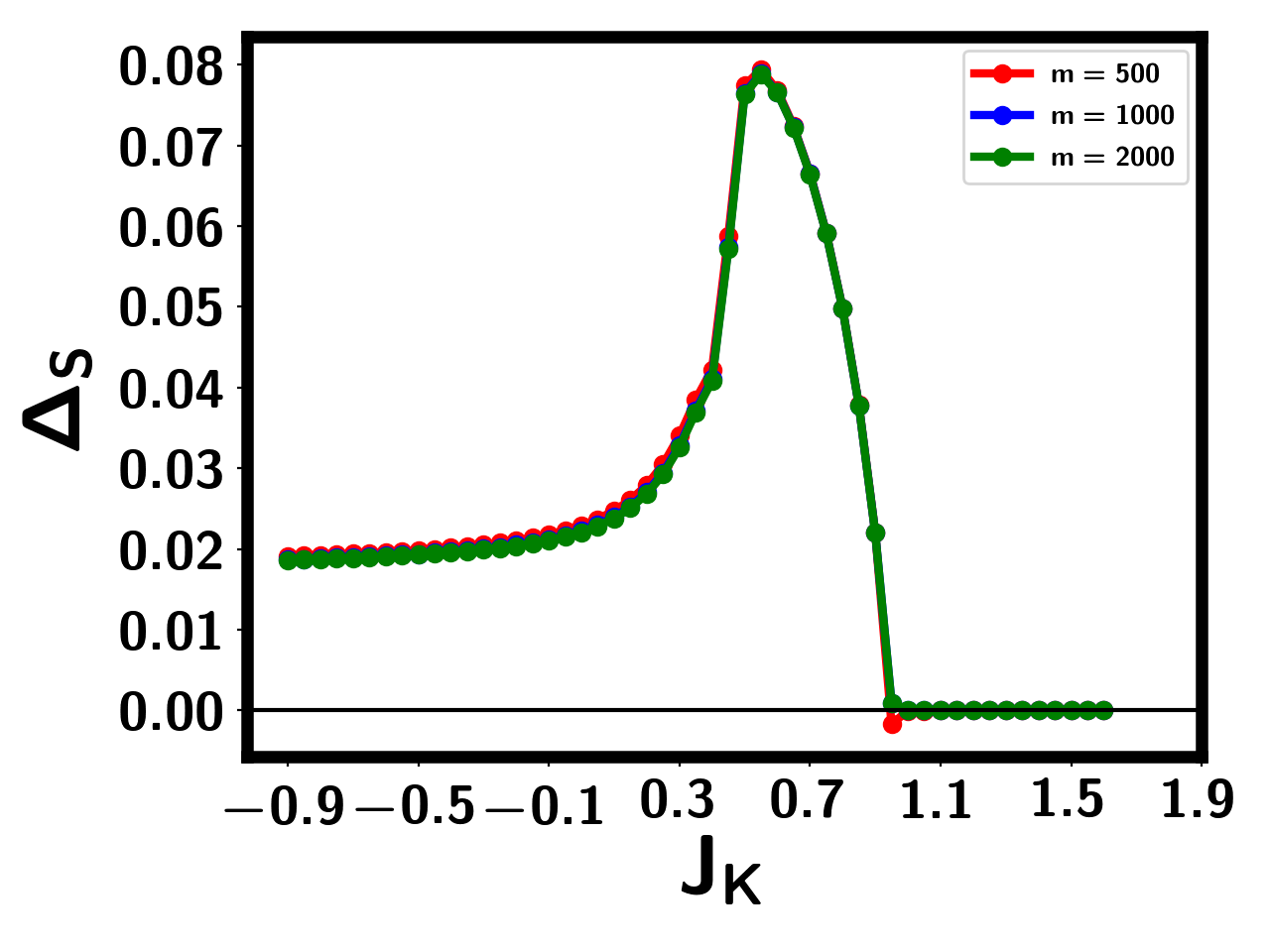}
\caption{Spin gap from finite DMRG with bond dimension $m=500,1000,2000$. We use $x=0.94$ with system size $L_x=100$. We use $t=1,J_c=J_s=0.5,J_{cs}=0.25$.}
\label{fig:spin_gap_bond_dimension}
\end{figure}

\begin{figure}[ht]
\centering
\includegraphics[width=0.85\textwidth]{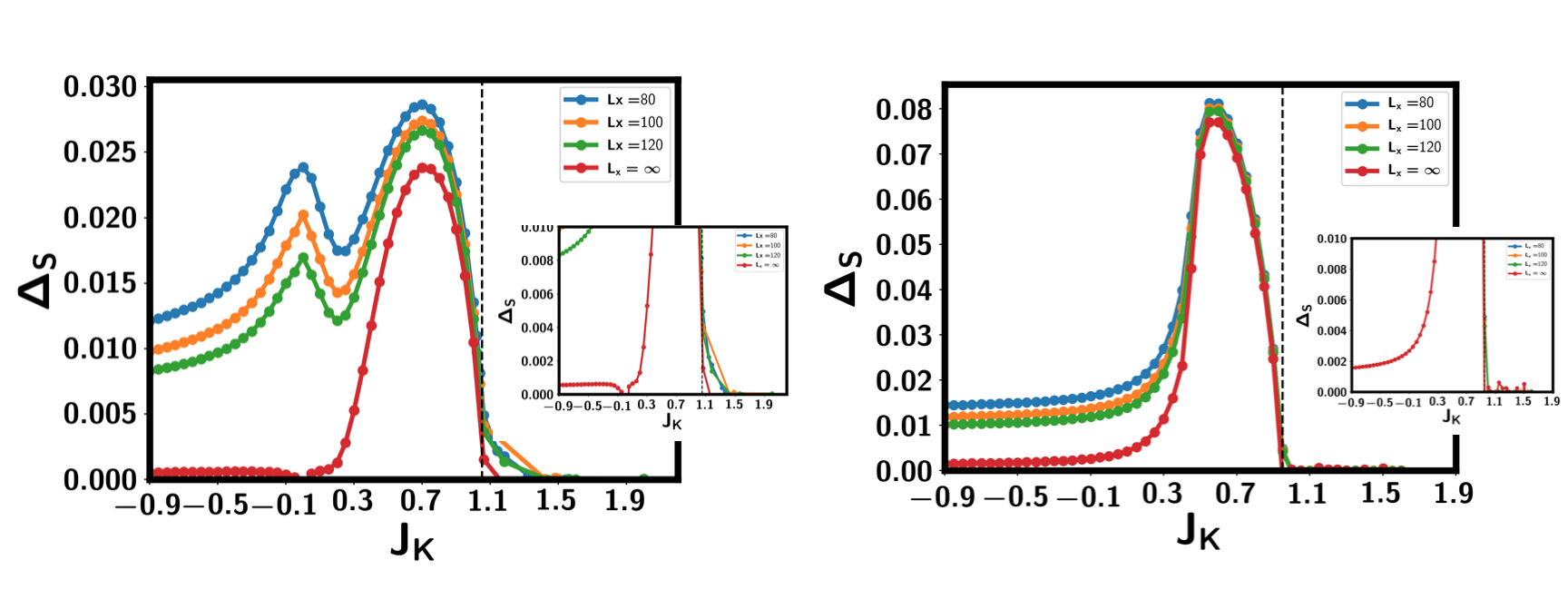}
\caption{Spin gap in the generalized Kondo model from finite DMRG at $x=0.9$ for (a)$J_{cs}=0$ and (b)$J_{cs}=0.2$. We use parameter $t=1,J_c=J_s=0.5$. The value at $L=\infty$ is extracted from polynomial fitting $\Delta_S(\frac{1}{L_x})=a \frac{1}{L_x^2}+b\frac{1}{L_x}+\Delta_S(L_x=\infty)$. In the inset we show a zoom in scale to demonstrate a finite spin gap at negative $J_K$ regime.}
\label{fig:spin_gap_L_scaling}
\end{figure}

\subsection{Spin gap and spin correlation length}

We report more results on spin gap from finite DMRG calculation of the generalized Kondo model. We always use $t=1,J_c=J_s=0.5$. We will vary $J_K,J_{cs}$ and the doping $x$. First, we show that the spin gap and our calculation converges when increasing  the bond dimension from $m=500$ to $m=2000$, shown in Fig.~\ref{fig:spin_gap_bond_dimension}. With bond dimension $m=2000$, we find that the truncation error is smaller than $10^{-7}$ inside the PDW phase and the energy convergence (difference between $m=1000$ and $m=2000$) is smaller than $10^{-6}$. We will use $m=2000$ in the remaining plots.

\begin{figure}[ht]
\centering
\includegraphics[width=0.85\textwidth]{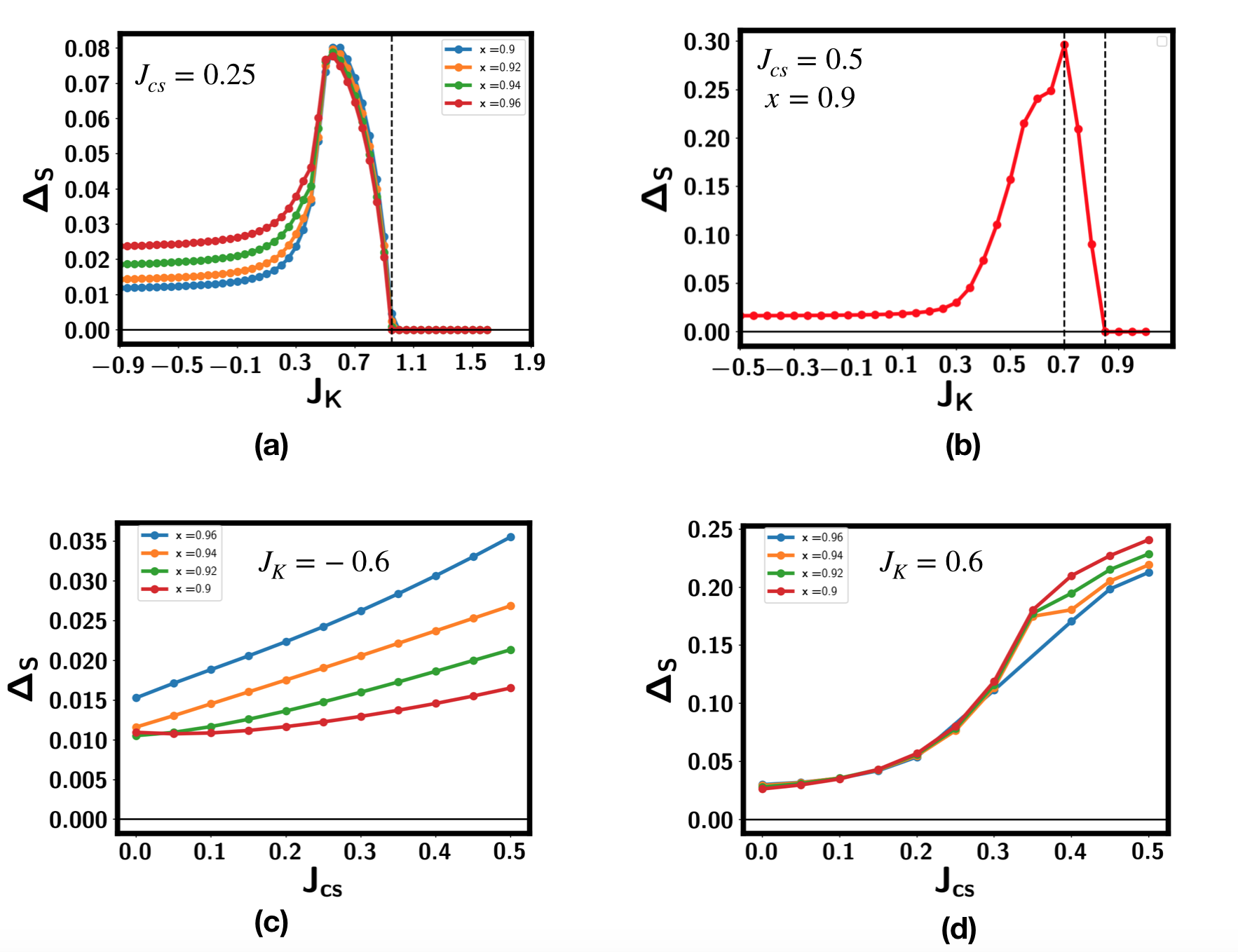}
\caption{(a) Spin gap with $J_K$ at $J_{sc}=0.25$, obtained with system size $L_x=100$. (b) Spin gap with $J_K$ at $J_{cs}=0.5$ and $x=0.9$. It is extrapolated to $L_x=\infty$ from data at $L_x=80,100,120$. (c) Spin gap with $J_{cs}$ at $J_K=-0.6$. (d) Spin gap with $J_{cs}$ at $J_K=0.6$. }
\label{fig:spin_gap_more}
\end{figure}

In Fig.~\ref{fig:spin_gap_L_scaling} we show how we extract the spin gap at $L=\infty$ from the results at finite $L=80,100,120$. One can see a finite spin gap when $J_K<J_K^c$, where $J_K^c=1.05, 0.95$ for $J_{cs}=0,0.25$ respectively. When $J_K>J_K^c$, we have a Luttinger liquid phase with zero spin gap.  We note that the spin gap is finite even at negative $J_K$ regime, though it is very small.  The spin gap at $J_K=-\infty$ can get enhanced if we increase $x$ or $J_{cs}$, as demonstrated in Fig.~\ref{fig:spin_gap_more}.

\begin{figure}[ht]
\centering
\includegraphics[width=0.85\textwidth]{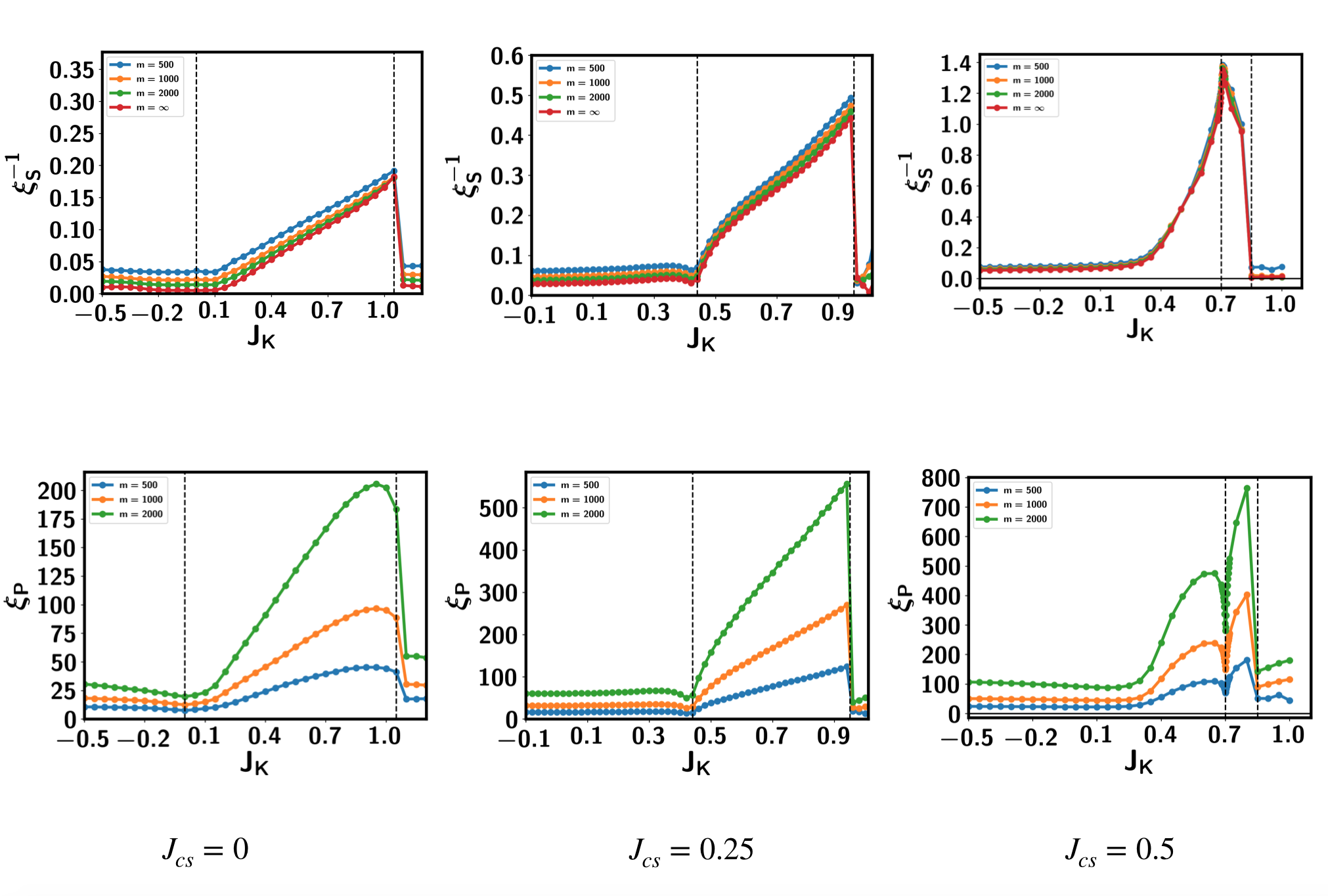}
\caption{Correlation lengths in infinite DMRG calculation at $x=\frac{28}{30}$ with unit cell size $L=30$. We use $t=1,J_c=J_s=0.5$. Correlation length is obtained from the transfer matrix method in one specific sector. The spin correlation length $\xi_S$ is from the sector $(\delta Q,\delta S_z)=(0,1)$. The typical operator in this sector is the $S^\dagger$ operator. The pairing correlation length $\xi_P$ is from the sector $(\delta Q,\delta S_z)=(2,0)$, with the typical operator as the Cooper pair operator. $m$ is the bond dimension. $\xi_S^{-1}(m=\infty)$ is extrapolated from the relation $\xi^{-1}_S(\frac{1}{m})=a \frac{1}{m^2}+b\frac{1}{m}+\xi^{-1}_S(m=\infty)$. (a)(d) $J_{cs}=0$.  (b)(e) $J_{cs}=0.25$. (c)(f) $J_{cs}=0.5$. The two dashed lines are at $J_K^0$ and $J_K^c$.}
\label{fig:idmrg_correlation_length}
\end{figure}

In addition to results from finite DMRG, we can also obtain correlation lengths using the transfer matrix techniques in infinite DMRG, as shown in Fig.~\ref{fig:idmrg_correlation_length}. We mainly care about the spin correlation length $\xi_S$, obtained in the sector with  $(\delta Q,\delta S_z)=(0,1)$and the pairing correlation length $\xi_P$, obtained in the sector with  $(\delta Q,\delta S_z)=(2,0)$. We show the data for $J_{cs}=0,0.25,0.5$. One can see that there is a finite $\xi_S^{-1}$ when $J_K<J_K^c$, consistent with a spin gap.  The pairing correlation length $\xi_P$ increases with the bond dimension $m$. At a fixed $m$, $\xi_P$ has a dip at $J_K^0$, where the Luttinger parameter $K_c$ also has a dip and the pairing correlation function power law decay exponent $K_{sc}=K^{-1}_c$ is peaked.  This is another evidence that there are two superconducting domes separated by $J_K^0$.

\subsection{Luttinger parameter $K_c$}

\begin{figure}[ht]
\centering
\includegraphics[width=0.8\textwidth]{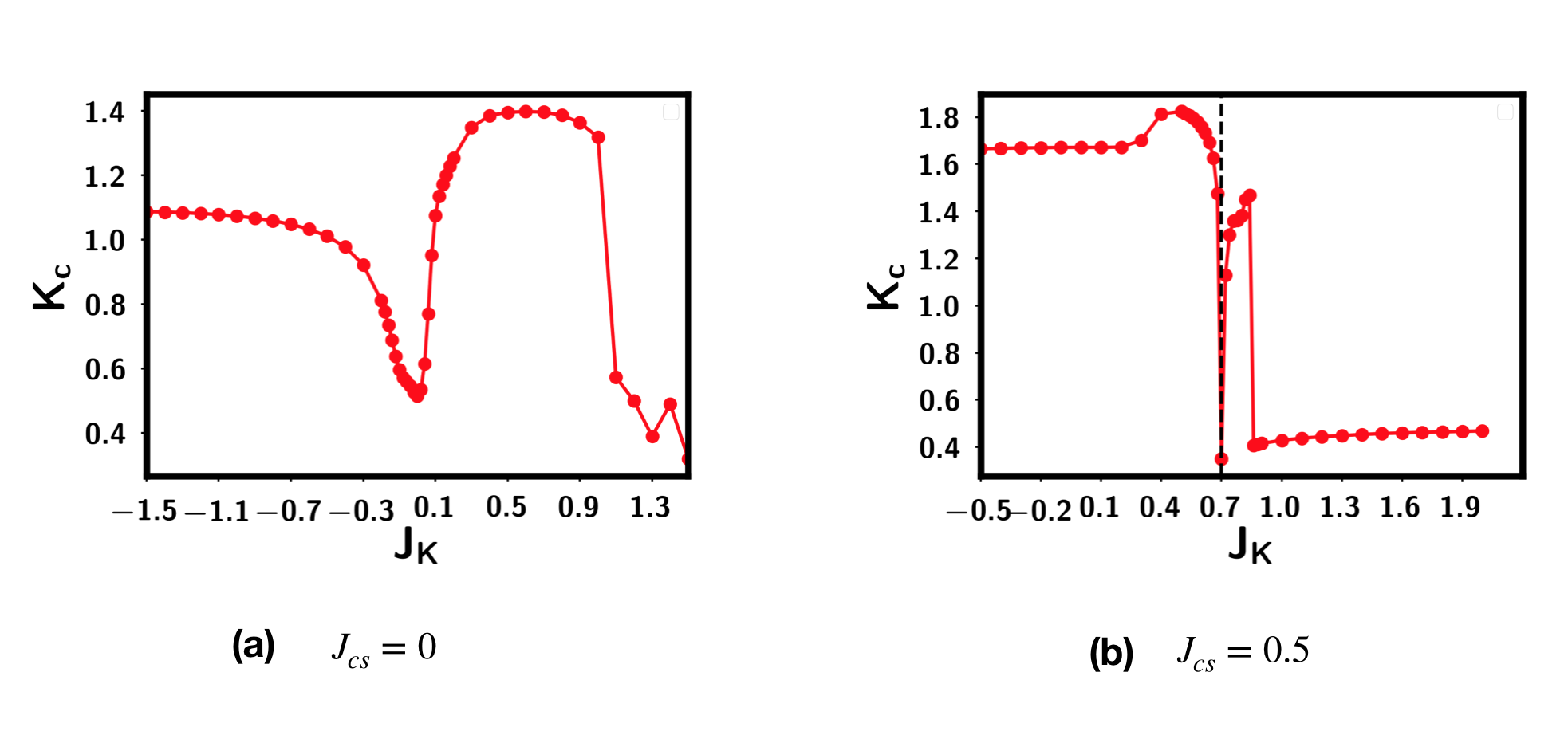}
\caption{Extracted  Luttinger parameter $K_c$ from finite DMRG at $L_x=100$ and $x=0.96$.(a)$J_{cs}=0$  (b)$J_{cs}=0.5$. The dashed line is at $J_K^0=0.7$ for $J_{cs}=0.5$.}
\label{fig:Kc_ksc}
\end{figure}

In Fig.~\ref{fig:Kc_ksc} we show the Luttinger parameter $K_c$ with $J_K$. The Luttinger parameter $K_c$ is fit using the method described in Sec.~\ref{appendix:extraction_luttinger_parameter}.   $K_c$ clearly has a dip at $J_K=J_K^0$,  The PDW phase is separated into two domes.  For the special case with $J_{cs}=0$, the $J_K^0=0$ point has zero spin gap and is a LL* phase with one charge mode and two spin modes. However, with a finite $J_{cs}$, the phase at $J_K^0$ is generically also  in the same Luther-Emery liquid phase with spin gap, although $K_c$ is smaller than one.

\subsection{Charge compressibility and Fermi velocity}

\begin{figure}[ht]
\centering
\includegraphics[width=0.8\textwidth]{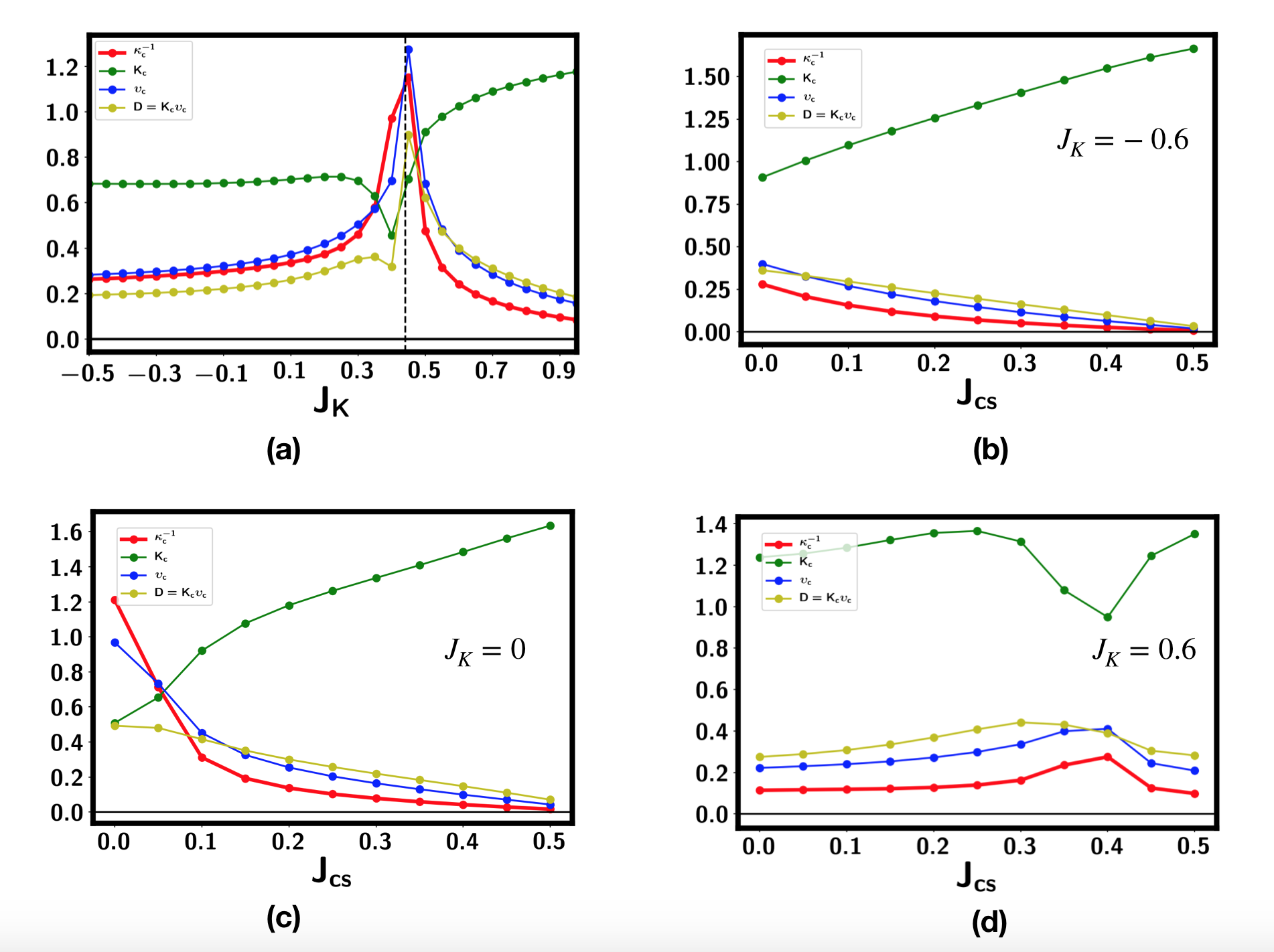}
\caption{Charge compressibility $\kappa_c$, luttinger parameter $K_c$, charge Fermi velocity $\upsilon_c$ and charge stiffness $D=K_c\upsilon_c$. (a)Finite DMRG at $x=0.94$ with system size $L_x=100$ and $J_{cs}=0.25$. (b)(c)(d) Change with $J_{cs}$ at fixed $J_{K}=-0.6,0,0.6$ for density $x=0.94$ with $L_x=100$.}
\label{fig:kappa_c_appendix}
\end{figure}

We also report the inverse charge compressibility $\kappa_c^{-1}$ and the Luttinger parameter $K_c$ in Fig.~\ref{fig:kappa_c_appendix}.  By using $\kappa_c=\frac{2 K_c}{\pi \upsilon_c}$, we can also obtain the Fermi velocity $\upsilon_c$ and the charge stiffness $D=K_c \upsilon_c$. We can see that there is a dip of $K_c$ and peak of the Fermi velocity $\upsilon_c$ at $J_K^0$, where the spin-spin correlation $\langle \vec{S}_{i;c}\cdot \vec{S}_i \rangle$  changes sign. Away from the $J_{K}^0$, $K_c$ gets enhanced and the Fermi velocity gets reduced, which is a signature of attractive interaction\cite{giamarchi2003quantum}.  If we stay in the FM or AF regimes with fixed sign of $\langle \vec{S}_{i;c}\cdot \vec{S}_i \rangle$, $J_{cs}$ term enhances $K_c$ and suppresses $\upsilon_c$, indicating stronger attraction. However, in Fig.~\ref{fig:kappa_c_appendix}(d), we find a dip of $K_c$ at $J_{cs}\approx 0.4$. This is again associated with the sign change of $\langle \vec{S}_{i;c}\cdot \vec{S}_i \rangle$ and vanishing of the polaron hybridization because we expect $J_K^0\approx 0.6$ when $J_{cs}=0.4$ based on the data in Fig.~\ref{fig:idmrg_correlation_length}.

\section{Spin polaron and its correlation functions}

In this section we show evidences for fermionic spin polaron at low energy and composite Cooper pair formed as bi-polarons in the generalized kondo model.

In the generalized Kondo model (Eq.~\ref{eq:kondo_t_J}), the fermionic spin polaron is defined as:

\begin{equation}
	\tilde c_{i;\sigma}=\frac{1}{2}(\vec{S}_i \cdot \vec{\sigma}_{\sigma \sigma'})c_{i;\sigma'}
\end{equation}
where $\vec S$ is the spin operator in the S layer and $c_{i;\sigma}$ is electron in the C layer. $\vec \sigma$ is the Pauli matrix which acts on the spin index of the electron operator.  It is easy to show that the spin polaron operator $\tilde c_{\sigma}$ has the same quantum number as the microscopic electron operator $c_{\sigma}$.

With the inter-layer spin-spin correlation, the polaron $\tilde c_{i;\sigma}$ will have finite overlap with the microscopic electron $c_{i;\sigma}$. Actually, one can find the hybridization to be

\begin{equation}
	\sum_{\sigma=\uparrow,\downarrow}c^\dagger_{i;\sigma}\tilde c_{i;\sigma}= \vec{S}_i \cdot \vec{S}_{i;c}
\end{equation}
which is nothing but the on-site  inter-layer spin-spin correlation. Here $\vec{S}_{i;c}=\frac{1}{2} c^\dagger_{i;\sigma}\vec{\sigma}_{\sigma \sigma'}c_{i;\sigma'}$ is the spin operator for the itinerant electron in the C layer.

 \subsection{Green functions of electron and polaron\label{appendix:sub_polaron_green}}

 We define electron-electron Green functions
 \begin{equation}
 	G_{\sigma;ee}(x,y)=\langle c^\dagger_\sigma(x) c_\sigma(y) \rangle
 \end{equation}

 Electron-polaron function:

 \begin{equation}
 	G_{\uparrow;ep}(x,y)= \langle c^\dagger_{\uparrow}(x) c_{\uparrow}(y)S_{z}(y) \rangle_c+\langle c^\dagger_{\uparrow}(x)c_{\downarrow}(y)S^{-}(y) \rangle_c
 \end{equation}
 and
 \begin{equation}
 	 G_{\downarrow;ep}(x,y)= -\langle c^\dagger_{\downarrow}(x)c_{\downarrow}(y)S_{z}(y) \rangle_c+\langle c^\dagger_{\downarrow}(x)c_{\uparrow}(y)S^{+}(y) \rangle
 \end{equation}

 Finally, polaron-polaron function:
 \begin{align}
 G_{\uparrow;pp}&=\langle c^\dagger_{\uparrow}(x)c_{\uparrow}(y)S_{z}(x)S_{z}(y) \rangle_c +\langle c^\dagger_{\downarrow}(x)c_{\downarrow}(y)S^\dagger(x)S^{-}(y)\rangle_c+\langle c^\dagger_{\uparrow}(x)c_{\downarrow}(y)S_z(x)S^{-}(y)\rangle_c+\langle c^\dagger_{\downarrow}(x)c_{\uparrow}(y) S^\dagger(x) S_{z}(y)\rangle_c \notag\\
 G_{\downarrow;pp}&=\langle c^\dagger_{\downarrow}(x)c_{\downarrow}(y)S_z(x)S_{z}(y)\rangle+\langle c^\dagger_{\uparrow}(x)c_{\uparrow}(y)S^{-}(x)S^{+}(y)\rangle-\langle c^\dagger_{\downarrow}(x)c_{\uparrow}(y)S_z(x)S^+(y)\rangle-\langle c^\dagger_{i;\uparrow}c_{j;\downarrow} S^{-}(x)S_{j;z}\rangle_c\notag\\
 \end{align}

 In the above $\vec{S}_i$ is the operator of the S layer and $c_{i;\sigma}$ is the operator in the C layer. $\langle O_C  O_S\rangle_c=\langle O_C O_S \rangle-\langle O_C \rangle \langle O_S \rangle$, where $O_C$ and $O_S$ are the operators in the C and S layer respectively.

 Inside the PDW phase, we find that
 \begin{equation}
 	G_{\alpha}(x)=A_{\alpha}e^{-\frac{x}{\xi_{\alpha}}}
 \end{equation}
 where $\alpha=$ee, ep, pp. 

 We always find that $\xi_{\alpha}$ is the same for all these three Green  functions, therefore we believe that the polaron and the electron both have overlaps with the same low energy mode. The amplitude $A_{\alpha}/A_{ee}$ for $\alpha=ep,pp$ thus are  characterizations of the mixture between the polaron and the electron.

 \subsection{Pairing-pairing correlation function for composite Cooper pair\label{appendix:sub_pair_pair}}

\begin{figure}[ht]
\centering
\includegraphics[width=0.8\textwidth]{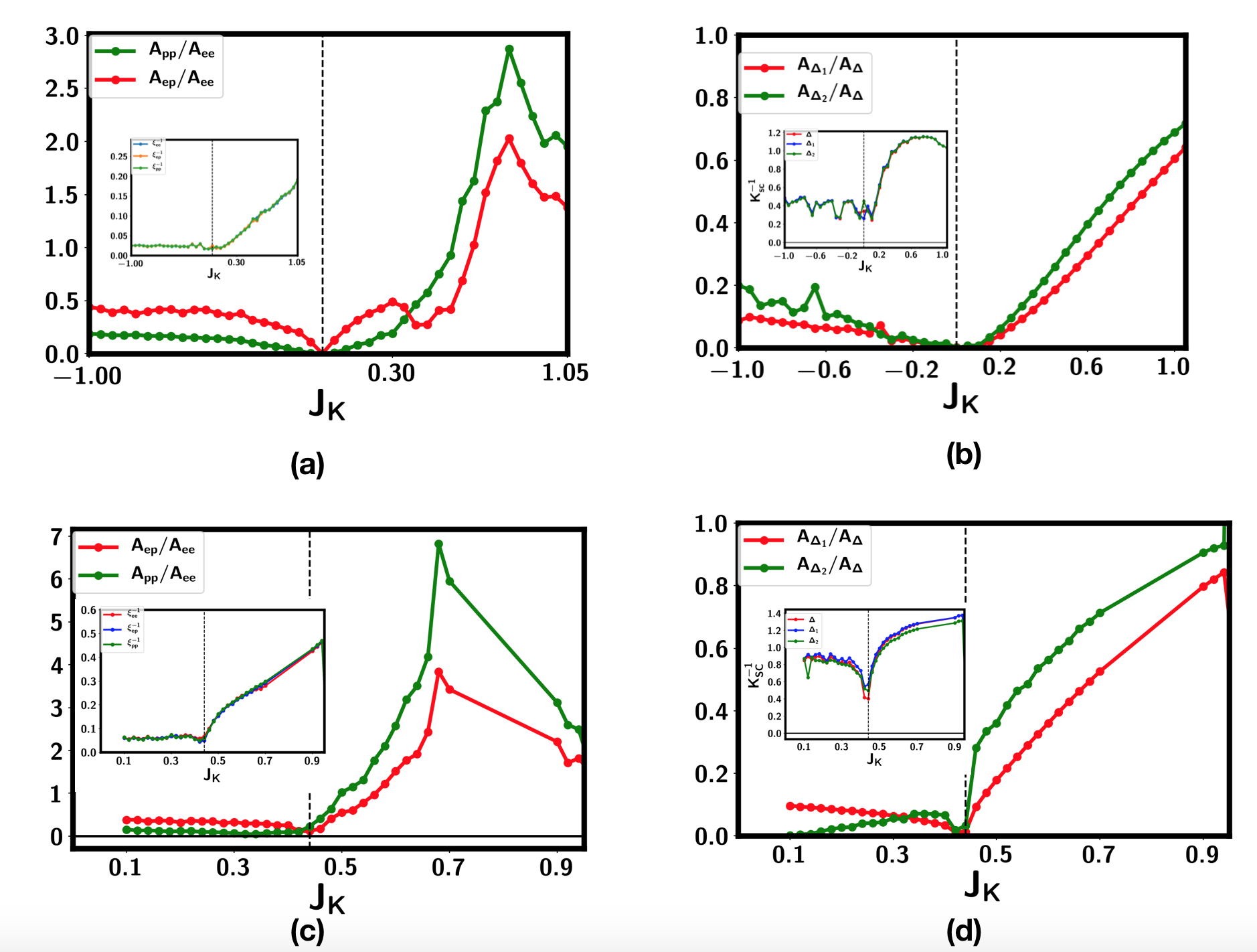}
\caption{Amplitudes and exponents for Green functions and pair-pair correlation functions obtained from infinite-DMRG with unit cell size $L=30$ and $x=\frac{28}{30}\approx 0.933$.  (a)(b) $J_{cs}=0$. The dashed line is  at $J_K^0=0$.  (c)(d)$J_{cs}=0.25$. The dashed line is at $J_K^0=0.44$.  (a)(c) are for Green functions define in Sec.~\ref{appendix:sub_polaron_green}. (b)(d) are for pairing correlation functions defined in Sec.~\ref{appendix:sub_pair_pair}. }
\label{fig:polaron_appendix_amplitude}
\end{figure}

With the spin polaron, it is easy to find that the spin-singlet pairing between electron and polaron:

\begin{equation}
	\epsilon_{\sigma \sigma'}(c_{i;\sigma}\tilde c_{j;\sigma'}+\tilde c_{i;\sigma} c_{j;\sigma})= \mathbf \Delta_{T;ij} \cdot (\mathbf S_i -\mathbf S_j)
\end{equation}
where the spin-triplet order parameters are:

\begin{equation}
	\mathbf \Delta_T=\big( c_{i;\downarrow}c_{j;\downarrow}-c_{i;\uparrow}c_{j;\uparrow}, -i (c_{i;\uparrow}c_{j;\uparrow}+c_{i;\downarrow}c_{j;\downarrow}), c_{i;\uparrow}c_{j;\downarrow}+c_{i;\downarrow}c_{j;\uparrow} \big)
\end{equation}

We can also define spin-singlet pairing  between  polarons:
 \begin{equation}
 	\epsilon_{\sigma \sigma'}\tilde c_{i;\sigma}\tilde c_{j;\sigma'}=-\Delta_{S;ij} (\mathbf S_i \cdot \mathbf S_j)+i\mathbf{\Delta_{T;ij}}\cdot (\mathbf S_i \times \mathbf S_j)
 \end{equation}

 If we define the Neel order parameter $\vec n(x)=(-1)^x(\vec{S}(x)-\vec{S}(x+1))$ and the VBS order parameter $\tilde V(x)=(-1)^x \vec{S}(x) \cdot \vec{S}(x+1)$ in the S layer,  we can see that the composite Cooper pairing order parameter $\vec{\Delta}_T\cdot \vec n$ can be understood as the Cooper pairing of one electron and one spin polaron and the composite pairing order $\Delta_S \tilde V$  is formed as a Cooper pair of spin polarons.

 Motivated by this observation,  in addition to the usual spin-singlet Cooper pair $\Delta(x)=\epsilon_{\sigma \sigma'}c_\sigma(x)c_{\sigma'}(x+1)$, we can define another two composite pairing order parameter: $\Delta_1(x)=\frac{1}{3} \vec \Delta_{T}(x) \cdot (\vec S(x) -\vec S(x+1))$ and $\Delta_2(x)=\Delta(x) \vec{S}(x) \cdot \vec{S}(x+1)$.  If we use spin rotation symmetry, we can further use $\Delta_1(x)=\Delta_{T;z}(x)(S_z(x)-S_z(x+1))$.

 Given that spin polaron is mixed with the single electron, we expect that $\Delta_1,\Delta_2$ are also mixed with the usual Cooper pair $\Delta$.  To characterize the mixture, we define the corresponding pairing-pairing correlation functions:

 \begin{equation}
 	\langle \Delta^\dagger_{\alpha}(x) \Delta_{\alpha}(0) \rangle_c=A_{\alpha}\frac{(-1)^x}{x^{K_{sc;\alpha}}}
 \end{equation}

 Again $\langle O(x) O(0) \rangle_c$ is defined by subtracting the connected part so  that it is zero at the decoupled limit $J_{cs}=J_K=0$ for $\alpha=\Delta_1,\Delta_2$.  Within the PDW phase, we find that $K_{sc}$ is the same for all these three correlation functions labeled by $\alpha=\Delta,\Delta_1,\Delta_2$, consistent with the expectation that they correspond to the same low energy mode.   The amplitudes $\frac{A_{\Delta_1}}{A_{\Delta}}$ and $\frac{A_{\Delta_2}}{A_{\Delta}}$ are then characterizations of the presence of the electron-polaron pair $\Delta_1$ and polaron-polaron pair $\Delta_2$.

\begin{figure}[ht]
\centering
\includegraphics[width=0.8\textwidth]{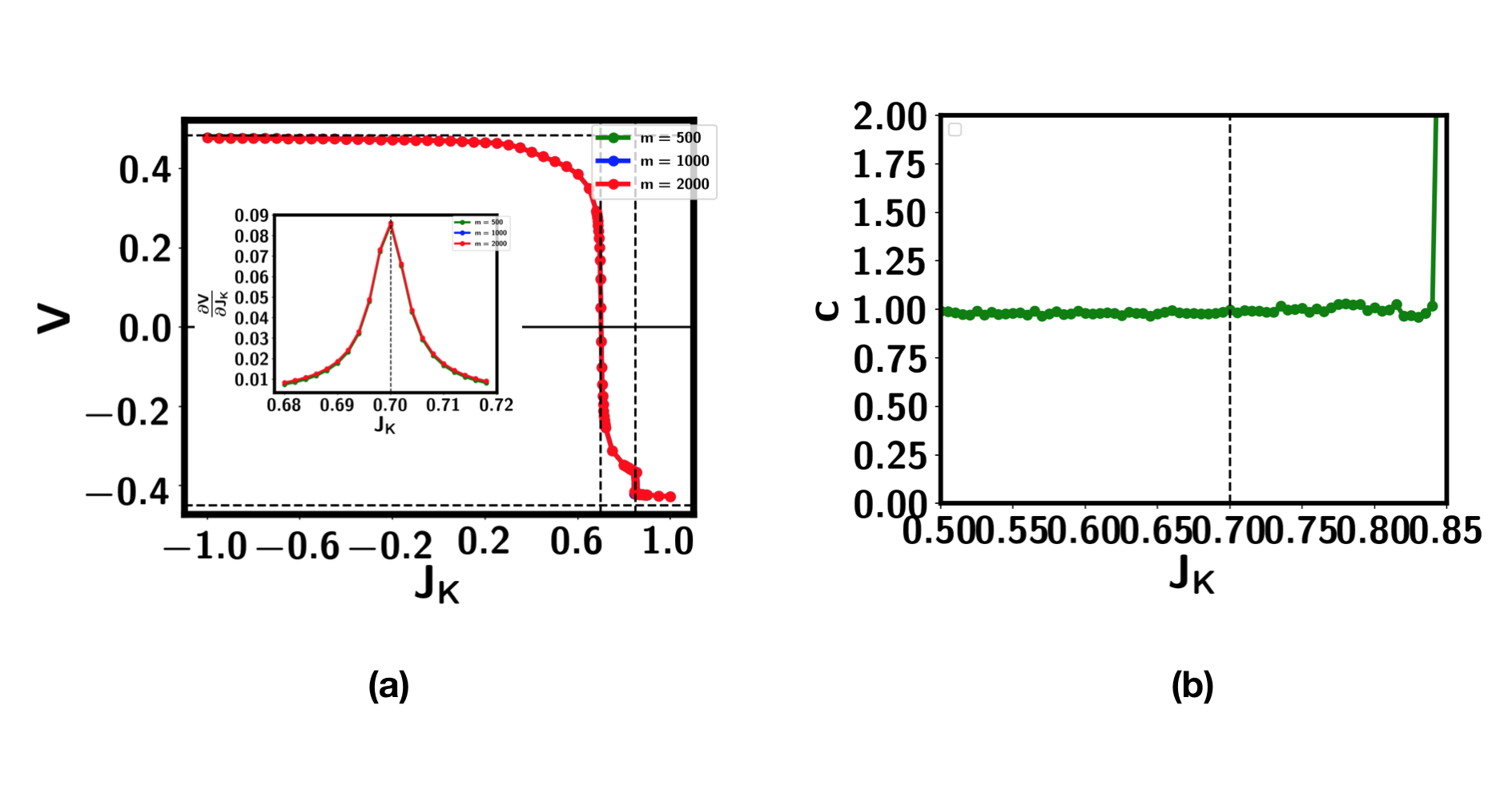}
\caption{(a) $\langle V_i \rangle$ for $J_{cs}=0.5$ from infinite DMRG. We use $x=\frac{14}{15}$ with unit cell size $L=30$. (b) Central charge $c$ fit from entanglement entropy $S=\frac{c}{6} \log \xi$, where $\xi$ is the correlation length. The two dashed lines are at $J_{K}^0=0.7$ and $J_K^c=0.85$. }
\label{fig:V_appendix}
\end{figure}

\subsection{Numerical results}

In Fig.~\ref{fig:polaron_appendix_amplitude} we show the amplitudes of the Green  function and the pairing correlation functions defined in the previous two subsections. We can see that the amplitude of the polaron Green  function and the polaron-polaron pairing correlation has a dip at $J_{K}^0$, where the Luttinger parameter $K_c$ also has a dip. This is an indication that the existences of the fermionic spin polaron and bipolarons are important to make $K_c<1$, which is required to get slow decay of the pairing correlation.

\section{Rapid crossover at $J_K^0$}

As shown in the previous sections, there is a dip of the Luttinger parameter $K_c$ at $J_K^0$, which separates the PDW phase into two regimes.  These two regimes have $\langle \vec S_{i;c} \cdot \vec S_i \rangle >0$ and $\langle \vec S_{i;c} \cdot \vec S_i \rangle <0$ separately. Here we address the question on whether there is a phase transition between the ferromagnetic and anti-ferromagnetic regimes of the PDW phase.  Our conclusion is that there is only a rapid crossover and there is no phase transition happening at $J_K^0$.  We have already shown the change of $\langle V_i \rangle= \langle \vec S_{i;c} \cdot \vec S_i \rangle+\frac{1}{4}$ for $J_{cs}=0.25$ in the main text.  In Fig.~\ref{fig:V_appendix} we show that for $J_{cs}=0.5$.  We can see that $V$ has a rapid change at $J_K^0=0.7$. $V$ is the first derivative of the energy with $J_K$, so the continuity of $V$ around $J_K^0$ rules out first order transition.  $\frac{\partial V}{\partial J_K}$ has a peak but does not diverge with the bond dimension $m$, ruling out a second order phase transition.  In Fig.~\ref{fig:V_appendix}(b), we can see that the central charge is $c=1$ across $J_K^0$, strongly suggesting that there is no true phase transition.

\begin{figure}[ht]
\centering
\includegraphics[width=0.8\textwidth]{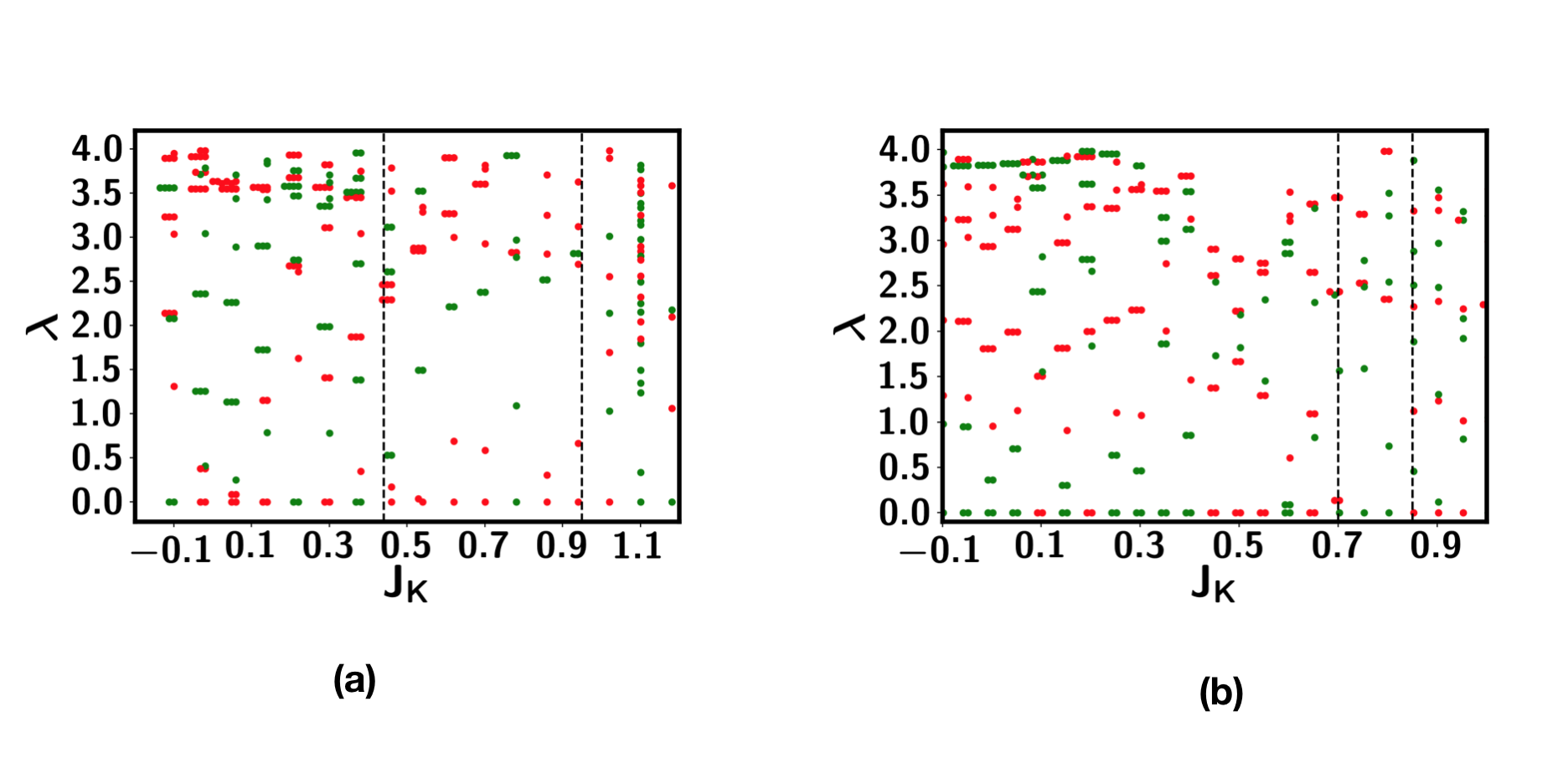}
\caption{Entanglement spectrum with $J_K$. (a) $J_{cs}=0.25$. The two dashed lines are at $J_K^0=0.44$ and $J_K^c=0.95$. (b) $J_{cs}=0.5$. The two dashed lines are at $J_K^0=0.7$ and $J_K^c=0.95$. }
\label{fig:ES_appendix}
\end{figure}

However, we find that there is a change in the entanglement spectrum at $J_K^0$, as shown in Fig.~\ref{fig:ES_appendix}.  Here for each $J_K$, we label the entanglement spectrum $\lambda_i$ with two colors for the two fermion parity.  When $J_K<J_K^c$, we can see that one fermion parity has even fold degeneracy, while the other one has odd fold degeneracy. Across $J_K^0$, the degeneracy of the lowest level changes from two fold to one fold. Despite this entanglement transition, we believe the system does not have a true phase transition. Also, we do not find any boundary states at zero energy with open boundary for $J_K<J_K^c$, consistent with the earlier discovery that the PDW is not topological\cite{may2020topology}.  Entanglement transition without boundary mode and true phase transition has also been reported in a different model\cite{Verresen}.

\section{More results on the Type II t-J model}

\begin{figure}[ht]
\centering
\includegraphics[width=0.95\textwidth]{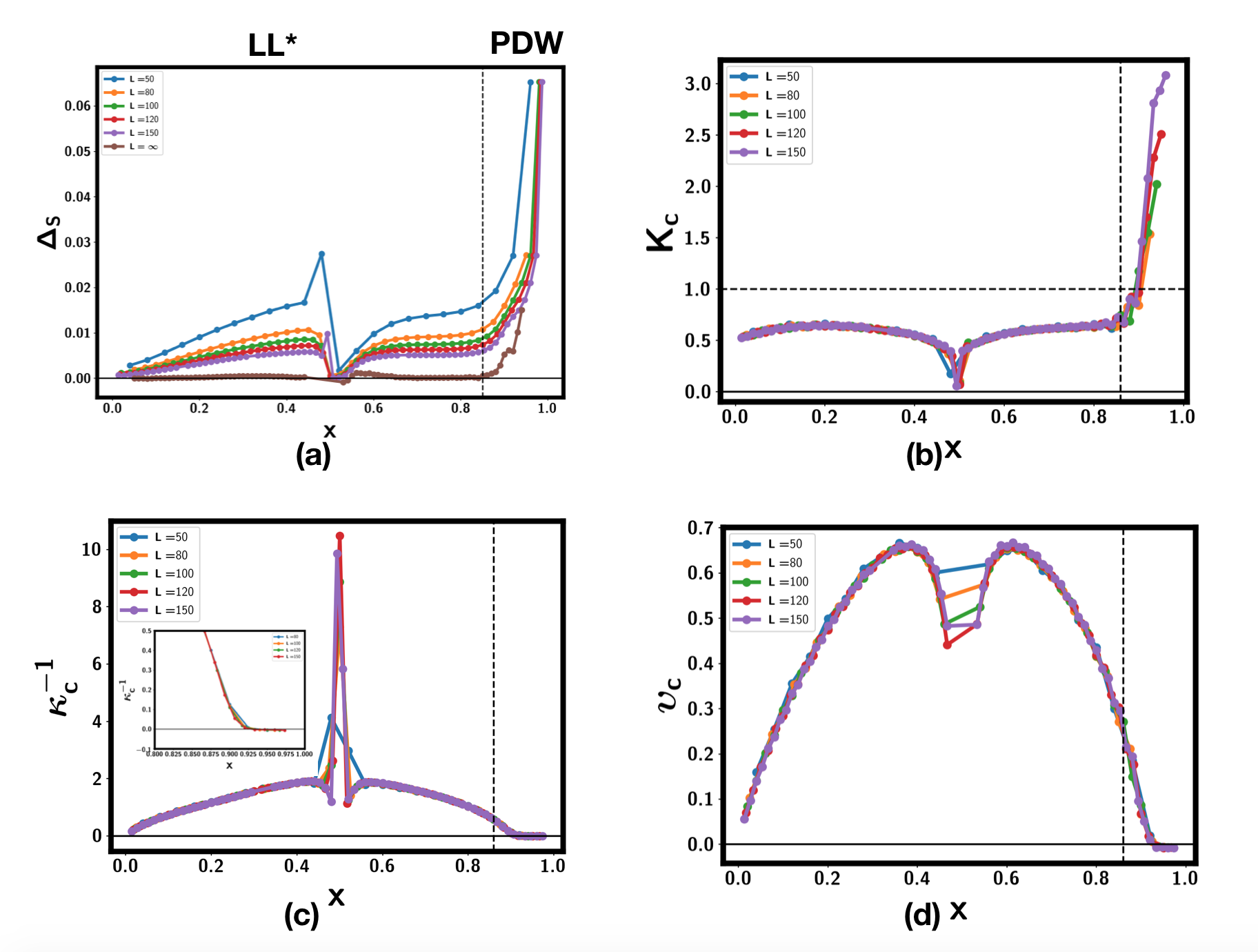}
\caption{Doping dependence in the type II t-J model with $t=1, J_s=J_d=0.5, J_{sd}=0.25$ from finite DMRG calculation. (a) Spin gap $\Delta_S$; (b)The Luttinger parameter $K_c$ for the charge mode; (c) The Charge compressibility $\kappa_c$; (d) The Fermi velocity of the charge mode $\upsilon_c$ fit from $\kappa_c=\frac{2K_c}{\pi \upsilon_c}$. $x=0.5$ is in a CDW phase and therefore there is discontinuity at this filling.}
\label{fig:1d_gap_compressibility}
\end{figure}

In this section we provide more data on the type II t-J model in one dimension.  In Fig.~\ref{fig:1d_gap_compressibility} we show the doping dependence for the spin gap, the Luttinger parameter $K_c$ and the charge compressibility for the parameter $t=1,J_s=J_d=0.5, J_{sd}=0.25$. We can see that there is an onset of the spin gap at $x_c=0.85$. When $x>0.85$, there is a quick increase of the Luttinger parameter $K_c$. We also find that the charge compressibility diverges when $x>0.93$.  This may suggest a phase separation phase, although the density profile in our finite DMRG calculation does not show phase separation and various correlation functions still look like a PDW phase.  The charge compressibility is $\kappa_c=\frac{2K_c}{\pi \upsilon_c}$ and a divergent $\kappa_c$ is associated with either a divergent of $K_c$ or vanishing of the velocity $\upsilon_c$.  We fit $\upsilon_c$ from $\kappa_c$ and $K_c$ and find it indeed vanishes after $x>0.93$. A rapid increase of $K_c$ and divergence of $\kappa_c$ has been also found in the conventional $t-J$ model in 1D when increasing $J/t$ to very large value\cite{ogata1991phase}. But there it needs $J/t>1$ which is unrealistic.  In our model, we find a rapid increase of $K_c$ even with realistic value of $J/t=0.5$. It may suggest that there is a much larger attractive interaction in the type II t-J model compared to the conventional t-J model.

\begin{figure}[ht]
\centering
\includegraphics[width=0.95\textwidth]{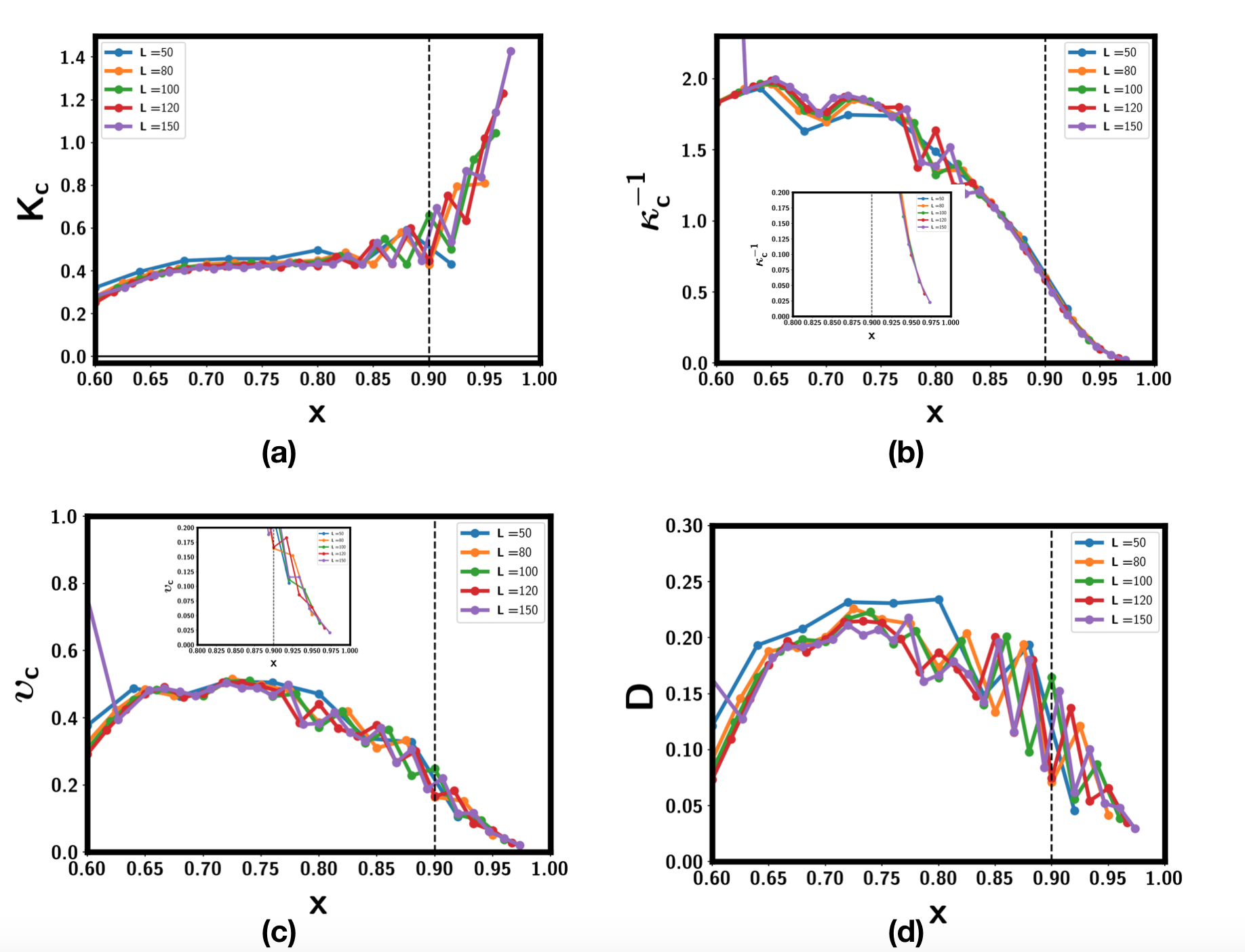}
\caption{Doping dependence in the type II t-J model with $t=1, J_s=J_d=0.3, J_{sd}=0.15$ from finite DMRG calculation. In this case we do not find the divergence of the charge compressibility in the density range we can reach.}
\label{fig:1d_gap_compressibility_smaller_J}
\end{figure}

One may question the existence of a stable PDW phase because the divergence of the charge compressibility would lead to a phase separated phase. Here we point out that a Luther-Emery liquid phase with a finite spin gap and finite charge compressibility exist in the range $0.85<x<0.93$ in Fig.~\ref{fig:1d_gap_compressibility}.  For this particular example it seems that $K_c<1$ in the region with finite $\kappa_c$.  However, this may be just a coincidence.  For example, if we use $t=1,J_s=J_d=0.3, J_{sd}=0.15$, we find that $\kappa_c$ is finite and $\upsilon_c>0$ for $x<0.975$ beyond which we do not have data. So there is no instability to phase separation for $x<0.975$, but we can still find $K_c>1$, as shown in Fig.~\ref{fig:1d_gap_compressibility_smaller_J}.   It is not clear whether a divergence of $\kappa_c$ will happen at larger $x$. In another word, whether we always have a phase separated regime between the PDW phase and the Haldane chain insulator at $x=1$ remains as an open question.

\begin{figure}[ht]
\centering
\includegraphics[width=0.85\textwidth]{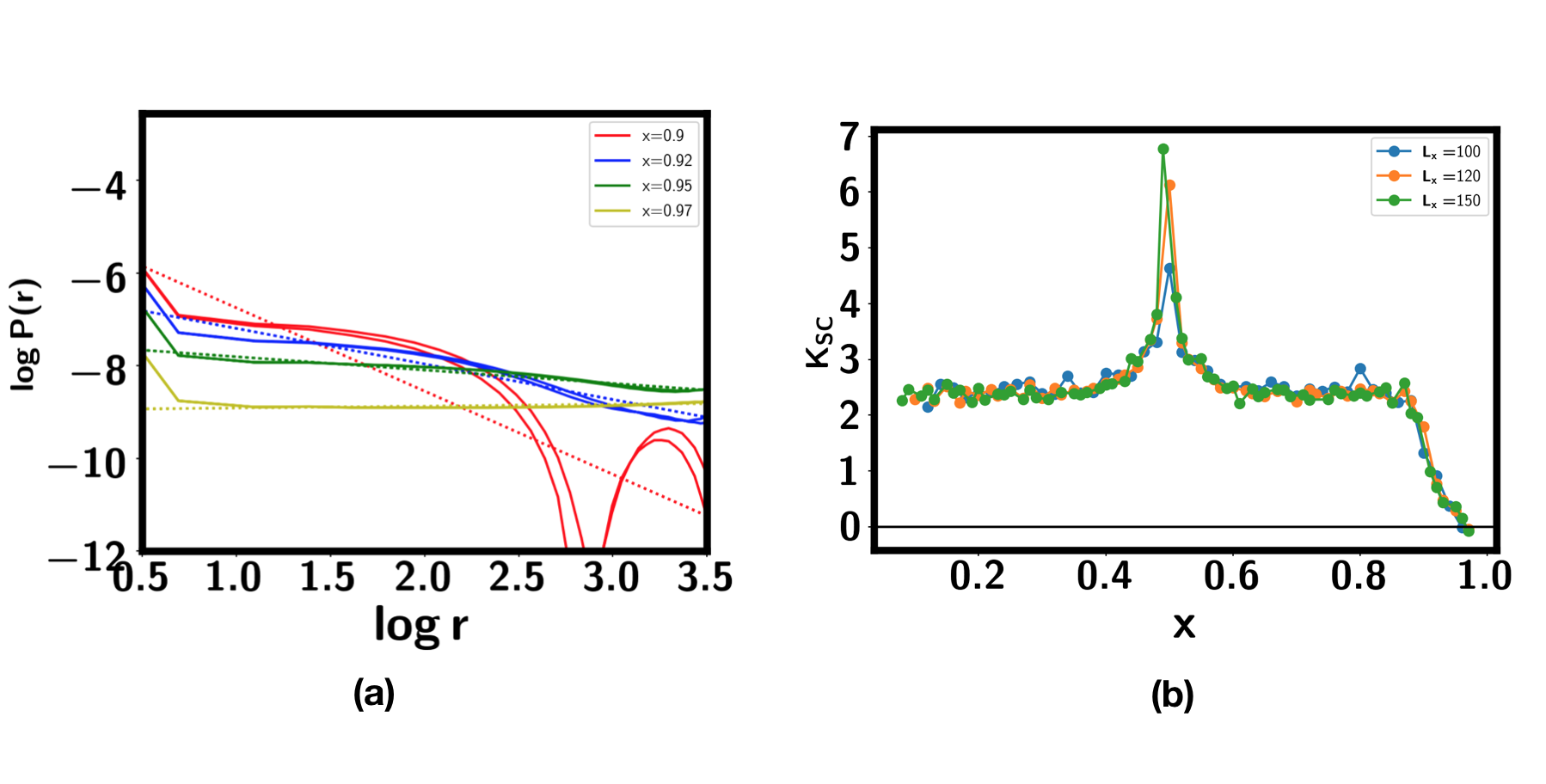}
\caption{(a)$\log |P(r)|$ vs $\log r$, where $P(r)=\langle \Delta_S(r) \Delta_S(0)\rangle$ is the pairing-pairing correlation function.  The dashed lines correspond to linear fit line whose exponent gives $K_{sc}$. (b)Doping dependence of the superconductor decaying exponent $K_{sc}$ for the type II t-J model. }
\label{fig:K_sc_type_II_appendix}
\end{figure}

A rapid increase of $K_c$ suggest a decrease of the exponent $K_{sc}$ corresponding to the algebraic decay of the pairing-pairing correlation function. We indeed find this behavior by explicitly fitting $K_{sc}$, shown in  Fig.~\ref{fig:K_sc_type_II_appendix} for $J_s=J_d=2J_{sd}=0.5$. In the regime $x<x_c$ (with $x_c \approx 0.85$), $K_{sc}$ is slightly larger than $2$, consistent with the result $K_{sc}=1+\frac{1}{K_c}$ with $K_c<1$ for a Luttinger liquid phase with repulsive interaction. However, when $x>x_c$, $K_{sc}$ quickly drops and approaches zero in the $x\rightarrow 1$ limit. This is consistent with the expectation $K_{sc}=\frac{1}{K_c}$ and the behavior of $K_c$ shown in Fig.~\ref{fig:1d_gap_compressibility}.

\begin{figure}[ht]
\centering
\includegraphics[width=0.85\textwidth]{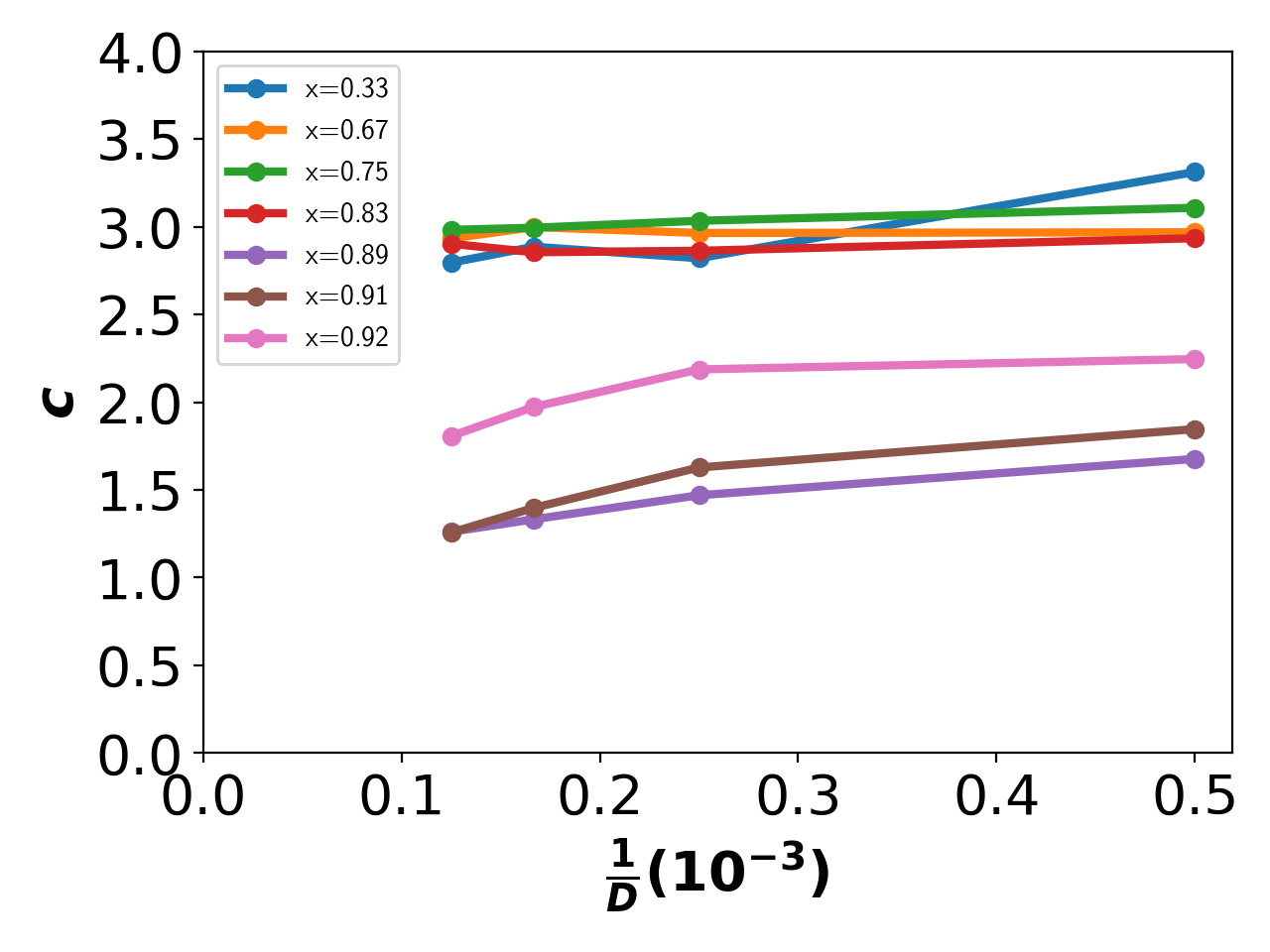}
\caption{Central charge at different doping $x$ in the type II $t-J$ model, using the same parameter as in Fig.~\ref{fig:1d_gap_compressibility}. The central charge is fit from the relation $S=\frac{c}{6} \log \xi$ in infinite DMRG, where $S$ is the entanglement entropy and $\xi$ is the correlation length.  Both $S$ and $\xi$ grow with the bond dimension $D$, therefore $c=6 \frac{\delta S}{\delta \xi}$ also changes with bond dimension $D$.  We plot $C$ with the bond dimension $D$ by varying $D$ from $2000$ to $8000$. For $x<0.85$, we find $c\approx 3$ in the whole range of $D$, consistent with a LL* phase with one charge mode and  two spin modes. For $x>0.85$, the central charge is smaller and decreases as $D$ increases. We believe $c$ will flow to $1$ in the $\frac{1}{D}\rightarrow 0$ limit.}
\label{fig:central_charge_type_II_1d}
\end{figure}

In the type II t-J model, clearly there is a phase transition between the LL* phase and the PDW phase at $x=0.85$ for $J_s=J_d=2J_{sd}=0.5$. The transition can also be visualized in the change of central charge as shown in  Fig.~\ref{fig:central_charge_type_II_1d}. We believe the transition is in the universality of Kosterlitz-Thouless transition, which will be discussed in Sec.~\ref{appendix:bosonization}.

\begin{figure}[ht]
\centering
\includegraphics[width=0.95\textwidth]{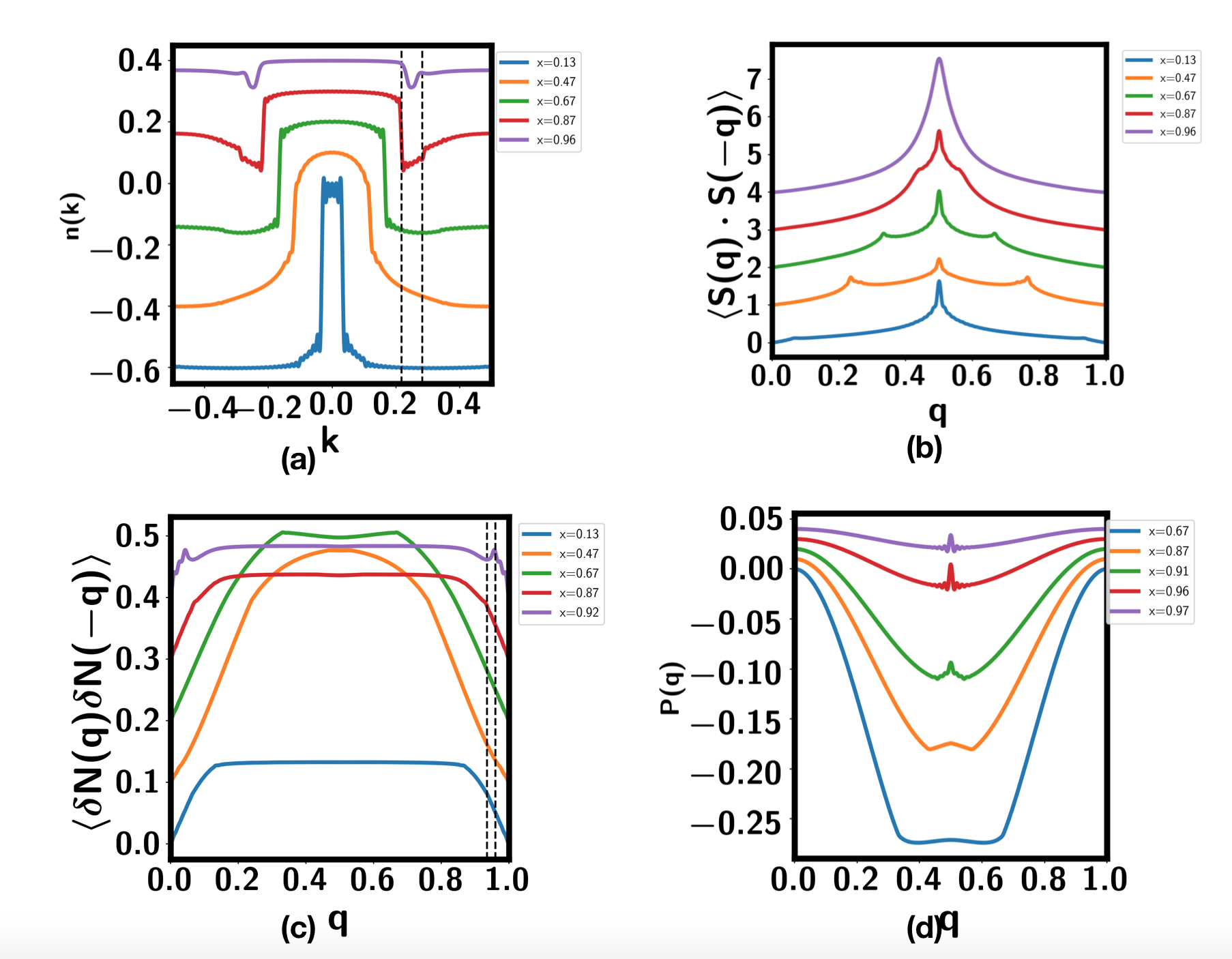}
\caption{Doping dependence of various correlation functions. (a)Momentum distribution function $n(k)=\sum_{\sigma}c^\dagger_{\sigma}(k)c_{\sigma}(k)\rangle$. There is a small Fermi surface with $k_F=\frac{x}{4} \times 2\pi$ which expands with $x$. When $x$ is large, there is also feature at $k_F+\pi$ from the scattering of the spin mode with $q=\pi$. The two dashed lines are at $k=k_F$ and $k=k_F+\pi$ at $x=0.85$. (b)Structure factor of the spin-spin correlation function. In the LL* phase when $x<0.85$, there are two modes at $q=2k_F=\frac{x}{2} \times 2\pi$ and at $q=\pi$. (c)Density-density structure factor. The two dashed lines correspond to $q=\frac{1+x}{2}\times 2\pi$ for $x=0.87$ and $x=0.92$. (d)$P(\mathbf q)=\langle \Delta_S(\mathbf q) \Delta_S(-\mathbf q) \rangle$, where $\Delta_S(x)=(c_{\uparrow}(x)c_{\downarrow}(x+1)-c_{\downarrow}(x)c_{\uparrow}(x+1))$ is the spin-singlet Cooper pair. Momentum is in units of $2\pi$.}
\label{fig:1d_cc_appendix}
\end{figure}

The PDW superconductor has a momentum $\mathbf q=\pi$, as shown in the peak at $\mathbf q=\pi$ of $P(\mathbf q)$ in Fig.~\ref{fig:1d_cc_appendix}(d). Here $P(\mathbf q)$ is the Fourier transformation of $P(r)$, which is the correlation function of spin-singlet Cooper pair between nearest neighbor sites. Inside the PDW phase, the density-density correlation function $\langle \delta N(q) \delta N(-q) \rangle$ shows peak at $q=\frac{1+x}{2}\times 2\pi$. In contrast, the peak is at either $2k_F$ or $4k_F$ with $2k_F=\frac{x}{2} \times 2\pi$ inside the LL* phase. This behavior can be captured in the bosonization theory provided in Sec.~\ref{appendix:bosonization}.

\section{Bosonization theory of PDW and its transition to LL* phase\label{appendix:bosonization}}

We provide a bosonization theory of the 1D PDW superconductor and its transition to the LL* phase. A very similar analysis has been performed for a two-leg Hubbard model in Ref.~\onlinecite{jaefari2012pair}.  For simplicity, we consider the generalized Kondo model in Eq.~\ref{eq:kondo_t_J}.  In the limit $J_H=J_{cs}=0$, the phase is apparently in a LL* phase with a conventional spinful Luttinger liquid decoupled plus a spin $1/2$ chain.

For the itinerant electron in the C layer, we have bosonization mapping:

\begin{equation}
  \psi_{r\sigma}=\frac{1}{\sqrt{2\pi\alpha}}U_{r,\sigma}e^{i r k_F x}e^{-\frac{i}{\sqrt{2}}\left(r \phi_c-\theta_c+\sigma(r \phi_s -\theta_s)\right)}
 \end{equation} 
 where $r=R,L$ labels the right-moving and the left-moving modes. $\sigma=\uparrow,\downarrow$ labels the spin. $U_{r,\sigma}$ is the Klein factor.

 Similarly, for the localized spin $1/2$ chain in the S layer, we have a spin mode $\tilde \theta_s, \tilde \phi_s$, while the corresponding charge mode $\tilde \theta_c,\tilde \phi_c$ is gapped ($\tilde \phi_c=0$) because it is in a Mott insulator.

 At $J_{cs}=0$, the Hamiltonian is

 \begin{align}
H&=\frac{\upsilon_c}{2\pi} \int dx K_c (\partial_x \theta_c)^2+\frac{1}{K_c} (\partial_x \phi_c)^2 \notag\\
&~~+\frac{\upsilon_s}{2\pi} \int dx  (\partial_x \theta_s)^2+ (\partial_x \phi_s)^2 \notag\\
&~~+\frac{\tilde \upsilon_s}{2\pi} \int dx  (\partial_x \tilde \theta_s)^2+(\partial_x \tilde \phi_s)^2
 \end{align} 
 where we have assumed that the Luttinger parameter for the spin modes are $K_s=\tilde K_s=1$ from the $SU(2)$ spin rotation symmetry.  $K_c$ is a function of $J/t$ and $x$. In the following we just treat it as a phenomenological parameter. Note that for spin $1/2$ chain in the S layer, we still use the convention that $K_s=1$ instead of $K=1/2$ as derived from fermionization of spin chain.

 Next, we need to add the $J_{cs}$ and $J_K=-J_H$ terms. To do that, we need to represent the electron spin $\mathbf{S}_{i;c}$ and the local spin $\mathbf{S}_i$ with the bosonization language. First, the spin of the C layer is:

 \begin{align}
S^z_c&=-\frac{1}{\sqrt{2}\pi} \partial_x \phi_s+\frac{1}{2\pi \alpha} \big( e^{-2i k_F x} e^{i \sqrt{2}\phi_c} \frac{1}{2}(\eta_1 \eta_3 e^{i\sqrt{2} \phi_s}-\eta_2 \eta_4 e^{-i \sqrt{2} \phi_s})+h.c.\big)\notag\\
S^+_c&=\frac{1}{2\pi \alpha}(\eta_1 \eta_4 e^{-i 2k_F x} e^{-i \sqrt{2}\theta_s} e^{i \sqrt{2} \phi_c}+\eta_3 \eta_2 e^{i 2k_F x} e^{-i \sqrt{2}\theta_s} e^{-i \sqrt{2} \phi_c}\notag\\
&+\eta_1 \eta_2 e^{-i \sqrt{2}\theta_s} e^{i \sqrt{2}\phi_s}+\eta_3\eta_4 e^{-i\sqrt{2}\theta_s}e^{-i\sqrt{2}\phi_s})
 \end{align}
 where $2k_F=\frac{x}{2} 2\pi$. $\eta_1,\eta_2,\eta_3,\eta_4$ are the Klein factors introduced to fix the fermion statistics.  We fix the gauge $\eta_1\eta_2\eta_3\eta_4=1$. For spin operators, the Klein factors can be suppressed by setting $\eta_1 \eta_3=\eta_2\eta_4=-i$, $\eta_1\eta_2=-\eta_3\eta_4=-i$ and $\eta_1\eta_4=\eta_3\eta_2=i$\footnote{Note that $\eta_a \eta_b=\pm i$ because $(\eta_a \eta_b)^2=1$. In the end we only care about the correlation function of the spin operators, where a pair of $\eta_a \eta_b$ appear. Thus whether $\eta_a \eta_b=i$ or $\eta_a\eta_b=-i$ does not matter. Here we make one choice just for simplicity.}.  Then we get

 \begin{align}
S^z_c&=-\frac{1}{\sqrt{2}\pi} \partial_x \phi_s+\frac{1}{2\pi \alpha} \sin \sqrt{2}\phi_s ( e^{-2i k_F x} e^{i \sqrt{2}\phi_c}+ e^{2i k_F x} e^{-i \sqrt{2}\phi_c})  \notag\\
S^+_c&=\frac{1}{2\pi \alpha}i( e^{-i 2k_F x} e^{-i \sqrt{2}\theta_s} e^{i \sqrt{2} \phi_c}+ e^{i 2k_F x} e^{-i \sqrt{2}\theta_s} e^{-i \sqrt{2} \phi_c})+e^{-i\sqrt{2}\theta_s}\sin \sqrt{2} \phi_s 
 \end{align}

For the local spin $\vec{S}_i$, we can use the same expression with spin mode $\tilde \theta_s, \tilde \phi_s$. This gives

 \begin{align}
\tilde S_z&=-\frac{1}{\sqrt{2}\pi} \partial_x \tilde \phi_s+(-1)^x \frac{1}{2\pi \alpha}  \sin \sqrt{2} \tilde \phi_s\notag\\
\tilde S^+&=\frac{1}{2\pi \alpha}((-1)^x ie^{-i \sqrt{2}\tilde \theta_s}+e^{-i \sqrt{2}\tilde \theta_s}\sin \sqrt{2}\tilde \phi_s)
 \end{align}

 Finally, we can write the inter-layer spin spin coupling as

 \begin{align}
H'&=\frac{g}{2\pi^2} \int dx \partial_x \phi_s \partial_x \tilde \phi_s-\frac{g}{8 \pi^2 \alpha^2} \int dx \cos 2 \theta_{s;-} \cos 2 \phi_{s;+}+\frac{g}{8 \pi^2 \alpha^2} \int dx \cos 2 \theta_{s;-} \cos 2 \phi_{s;-}
\end{align}
 where $g=2 J_{cs}-J_H$ and $g'=2 J_{cs}-J_H$. We defined $\theta_{s;\pm}=\frac{1}{\sqrt{2}}(\theta_s\pm \tilde \theta_s)$ and $\phi_{s;\pm}=\frac{1}{\sqrt{2}}(\phi_s \pm \tilde \phi_s)$.

 The first line will renormalize the Luttinger parameter $K_\pm$ for the $\theta_{s;\pm}$ mode. For simplicity we assume $\upsilon_s=\tilde \upsilon_s=\upsilon$ at the initial point. We will have:

 \begin{align}
H&=\frac{\upsilon_c}{2\pi} \int dx K_c (\partial_x \theta_c)^2+\frac{1}{K_c} (\partial_x \phi_c)^2 +\frac{\upsilon_+}{2\pi} \int dx  K_{+}(\partial_x \theta_{s;+})^2+ \frac{1}{K_{+}}(\partial_x \phi_{s;+})^2 +\frac{\upsilon_-}{2\pi} \int dx K_{-} (\partial_x \theta_{s;-})^2+\frac{1}{K_{-}}(\partial_x \phi_{s;-})^2 \notag\\
&~~-\frac{g}{8 \pi^2 \alpha^2} \int dx \cos 2 \theta_{s;-} \cos 2 \phi_{s;+}+\frac{g'}{8 \pi^2 \alpha^2} \int dx \cos 2 \theta_{s;-} \cos 2 \phi_{s;-}
\end{align}
where,
\begin{align}
\upsilon_\pm &= \upsilon \sqrt{1\pm \frac{g}{2\pi \upsilon}}\notag\\
K_\pm &=\frac{1}{\sqrt{1\pm \frac{g}{2\pi \upsilon}}}
\end{align}

Note that in the above we have ignored the intra-layer spin-spin coupling. We assume that at the decoupled limit the intra-layer super-exchange terms $J_c,J_s$ are not strong enough to destroy the LL* phase. Here we are mainly interested in the possible instability of the LL* phase caused by the inter-layer spin-spin coupling terms $J_H$ and $J_{cs}$.  The scaling dimension of $g'$ is $[g']=2-(K_-+\frac{1}{K_-})< 0$ and the $g'$ term is generically irrelevant.   In the following we only keep the $g$ term. The RG equation is 

\begin{align}
\frac{d K_+}{dl}&=-\frac{1}{4}K_+^2 g^2\notag\\
\frac{d K_-}{dl}&=\frac{1}{4}g^2 \notag\\
\frac{d g}{dl}&=(2-K_{+}-\frac{1}{K_-}) g
\end{align}

We note that if $g>0$ initially, then $K_{+}+\frac{1}{K_-}<2$ and $g$ flows to $+\infty$. The $g$ term will pin $\theta_{s;-}$ and $\phi_{s;+}$, so both spin modes are gapped out and we are left with only the charge mode $\theta_c,\phi_c$.  This turns out to be a PDW superconducting phase which we will describe later.  On the other hand, if $g<0$, then $K_{+}+\frac{1}{K_-}>2$ initially and $g$ flows to zero, while $K_{+},K_{-}$ flow to $1$, resulting in the LL* phase. Therefore changing $g$ tunes a phase transition between the LL* phase and the PDW phase.

 We note that the oscillating part of the spin operator in the C layer does not enter the final Hamiltonian because $2k_F=(1-x) \pi$ is incommensurate and can not cancel the $\mathbf q=\pi$ part of the $\tilde S$.  In contrast, for the $x=1$ point, the oscillatory part also enters the Hamiltonian and there is term like $-g_1 \cos 2\phi_{s;+}-g_2 \cos 2\phi_{s;-}$, which is more relevant than the $g$ and $g'$ term. Therefore the above analysis only works for $x<1$ and will break down for the filling $x=1$, where we will get Mott insulator with spin in either Haldane phase or a rung singlet phase.

 The above analysis suggests that there is a Kosterlitz-Thouless (KT) transition between a LL* phase ($g<0$) and a PDW phase ($g>0$). The central charge changes from $c=3$ to $c=1$. The same transition has been found at $x_c=0.85$ of the type II t-J model shown in Fig.~\ref{fig:1d_gap_compressibility} and Fig.~\ref{fig:central_charge_type_II_1d}. In the following we study the property of the PDW phase in details based on bosonization language.

 \subsection{Property and Order parameter of the PDW phase}

 We discuss the property of the PDW superconductor phase in the $g>0$ region..  The $- g\cos 2 \theta_{s;-} \cos 2 \phi_{s;+}$ will pin $\theta_{s;-}$ and $\phi_{s;+}$ into either $\theta_{s;-}=\phi_{s;+}=0$ or $\theta_{s;-}=\phi_{s;+}=\frac{\pi}{2}$. Next we will study various correlation functions and show that this is a PDW phase with a composite pairing operator.

 Because only the charge mode survives, the phase must be a Luther-Emery liquid with a gap for spin and single electron excitation. The only order parameter we need to consider is the pairing order and the charge-density-wave (CDW) orders. Here we will show that the pairing and CDW order within the C layer is actually also gapped in the sense that its correlation function is exponentially decayed. The only gapless order parameter is a composite object by combining the order parameter in the C layer with the Neel or valence-bond-solid (VBS) order parameter in the S layer.

 The zero-momentum spin-singlet superconductor order parameter within C layer is 
\begin{equation}
  \Delta_S=\psi_{R\uparrow}\psi_{L\downarrow}-\psi_{R\downarrow}\psi_{L\uparrow}=\frac{i}{\pi \alpha}e^{i \sqrt{2}\theta_c} \cos \sqrt{2} \phi_s
\end{equation}

also  zero-momentum spin-triplet pairing within C layer is

\begin{align}
\mathbf \Delta_T&=\big( \psi_{R\downarrow}\psi_{L\downarrow}-\psi_{R\uparrow}\psi_{L\uparrow}, -i (\psi_{R\uparrow}\psi_{L\uparrow}+\psi_{R\downarrow}\psi_{L\downarrow}), \psi_{R\uparrow}\psi_{L\downarrow}+\psi_{R\downarrow}\psi_{L\uparrow} \big)=\frac{1}{\pi \alpha}e^{i\sqrt{2} \theta_c}(\sin \sqrt{2}\theta_s, \cos \sqrt{2}\theta_s,\sin \sqrt{2}\phi_s)
\end{align}

We can write down density operator as:
\begin{equation}
  \rho(x)=-\frac{\sqrt{2}}{\pi} \partial_x \phi_c+e^{i 2 k_F x} \rho_{2k_F}(x)+e^{-i 2 k_F x} \rho_{-2k_F}
\end{equation}
where the CDW order at momentum $Q=2k_F=2\pi \frac{x}{2}$ is
\begin{equation}
  \rho_{2k_F}(x)=\frac{i}{\pi \alpha} e^{-i \sqrt{2} \phi_c}\cos(\sqrt{2} \phi_s)
\end{equation}

All of these order parameters contain terms like $\cos  \sqrt{2} \phi_s$, $\cos \sqrt{2}\theta_s$, $\sin \sqrt{2}\phi_s$, $\sin \sqrt{2} \theta_s$. Because $\phi_s=\frac{1}{\sqrt{2}}(\phi_{s;+}+\phi_{s;-})$ and $\theta_s=\frac{1}{\sqrt{2}}(\theta_{s;+}+\theta_{s;-})$, the correlation functions of these terms are exponentially decayed. This is because we can not pin $\phi_{s;+},\phi_{s;-}$ at the same time. For example, let us consider $\cos \sqrt{2}\phi_s=\cos (\phi_{s;+}+\phi_{s;-})$. In the PDW phase, $\phi_{s;+}$ and $\theta_{s;-}$ is pinned, so $\phi_{s;-}$ is gapped. $\cos (\phi_{s;+}+\phi_{s;-})\sim \cos \phi_{s;-}$ has exponentially decayed correlation function.

In the following we show that certain composite order parameter still has power law correlation function. The key idea is to cancel the factor like $\cos (\phi_{s;+}+\phi_{s;-})$ by combining an order parameter from the S layer.  First, in S layer we can define Neel order parameter through $\vec{S}=\vec{S}_0+(-1)^x \vec n$.  Here $\vec n$ is the Neel order parameter with a momentum $Q=\pi$. It is easy to find that $(n_x,n_y,n_z)=\frac{1}{2\pi \alpha}(\sin \sqrt{2}\tilde \theta_s, \cos \sqrt{2}\tilde \theta_s,\sin \sqrt{2}\tilde \phi_s)$. Meanwhile, there is a VBS order parameter defined through $\vec{S}_i \cdot \vec{S}_j \sim (-1)^i \tilde V$. The VBS order parameter V carries momentum $\mathbf Q=\pi$ and can be expressed as $\tilde V=\frac{1}{2\pi \alpha} \cos \sqrt{2}\tilde \phi_s$.

Now we can define a composite order parameter $\Delta_{\text{PDW}}  \sim \Delta_S \tilde V \sim e^{i\sqrt{2} \theta_c} (\cos 2\phi_{s;+}+\cos 2\phi_{s;-})$. We note that $\phi_{s;+}$ is pinned while $\phi_{s;-}$ is fluctuating, so we only needs to keep $\cos 2\phi_{s;+}$ term which is basically a constant. In the end we find $\Delta_S \tilde V\sim e^{-i \sqrt{2} \theta_c}$.  Actually, we can also find that $\vec{\Delta}_T \cdot \vec n \sim e^{-i\sqrt{2} \theta_c}$.  Therefore, we have the composite PDW order parameter:

\begin{equation}
	O_{\text{PDW}}\sim \Delta_S \tilde V \sim \vec{\Delta}_T \cdot \vec n  \sim e^{-i \sqrt{2} \theta_c}
\end{equation}
which carries momentum $Q_{\text{PDW}}=\pi$ and is spin singlet.

Similarly, one can define a composite CDW order parameter 
\begin{equation}
  O_{CDW}\sim \rho_{2k_F}(x) \tilde V\sim e^{-i\sqrt{2} \phi_c}
\end{equation}
which carries momentum $Q_{\text{CDW}}=2k_F+\pi=\frac{1+x}{2} 2\pi$.

It is easy to find correlation function of the PDW and CDW order parameters: 

\begin{equation}
  O_{\text{PDW}}(x) O_{\text{PDW}}(0) \sim \frac{1}{x^{\frac{1}{K_c}}}
\end{equation}

and

 \begin{equation}
  O_{\text{CDW}}(x) O_{\text{CDW}}(0) \sim \frac{1}{x^{K_c}}
\end{equation}

One can see that both the PDW and CDW order parameter have power law decay correlation functions, though their exponents are inverse to each other. This is a typical behavior of Luther-Emery liquid.  When $x$ is close to $1$, we find $K_c>1$ in our DMRG calculation, thus PDW order dominates over the CDW order. This is the reason why we call the phase as PDW superconductor.

\end{document}